\documentclass[onecolumn]{aastex631}
\usepackage{color}
\usepackage[caption=false]{subfig}
\newcommand\lya{Ly$\alpha$}
\usepackage{rotating}
\shorttitle{Constraining Reionization with VPF of Roman LAEs}
\shortauthors{Perez et al.}
\accepted{March 23, 2022}
\submitjournal{ApJ}
\begin{document}

% \title{Evolution of the Void Probability Function of Lyman-Alpha Emitters \\ throughout Reionization, as observed with the Roman Space Telescope}
\title{Constraints on the Epoch of Reionization with Roman Space Telescope \\ and the Void Probability Function of Lyman-Alpha Emitters}

\correspondingauthor{Lucia A. Perez}
% \email{lucia.perez@asu.edu}
\email{lucia.perez.phd@gmail.com}

\author[0000-0002-8449-1956]{Lucia A. Perez}
\affiliation{Department of Astrophysical Sciences,
Princeton University,
4 Ivy Lane,
Princeton, NJ 08544, USA}
\affiliation{Arizona State University,
School of Earth and Space Exploration,
781 Terrace Mall,
Tempe, AZ 85287, USA}

\author[0000-0002-9226-5350]{Sangeeta Malhotra}
\affiliation{NASA Goddard Space Flight Center,
8800 Greenbelt Road,
Greenbelt, MD 20771, USA}

\author[0000-0002-1501-454X]{James E. Rhoads}
\affiliation{NASA Goddard Space Flight Center,
8800 Greenbelt Road,
Greenbelt, MD 20771, USA}

\author[0000-0002-0784-1852]{Isak G.B. Wold}
\affiliation{NASA Goddard Space Flight Center,
8800 Greenbelt Road,
Greenbelt, MD 20771, USA}
% \affil{Astrophysics Science Division, Goddard Space Flight Center, Greenbelt, MD 20771, USA}
\affil{Department of Physics, The Catholic University of America, Washington, DC 20064, USA }
\affil{Center for Research and Exploration in Space Science and Technology, NASA/GSFC, Greenbelt, MD 20771, USA}

\begin{abstract}
% \textbf{\textit{NEW: WITH VOLUME-AVERAGED FRACTIONS FOR JENSEN!}}

We use large simulations of Lyman-Alpha Emitters with different fractions of ionized intergalactic medium to quantify the clustering of Ly$\alpha$ emitters as measured by the Void Probability function (VPF), and how it evolves under different ionization scenarios. We quantify how well we might be able to distinguish between these scenarios with a deep spectroscopic survey using the future Nancy Grace Roman Space Telescope. Since Roman will be able to carry out blind spectroscopic surveys of Ly$\alpha$ emitters continuously between $7<z<12$ to sensitivities of at least
% a sensitivity of approximately 
$10^{-17}$ erg sec$^{-1}$ over a wide field of view, it can measure the epoch of reionization as well as the pace of ionization of the intergalactic medium (IGM). We compare deep Roman surveys covering roughly 1, 4, and 16 deg$^2$, and quantify what constraints on reionization the VPF may find for these surveys.
A survey of 1 deg$^2$ would distinguish between very late reionization and early reionization to 3$\sigma$ near $z=7.7$ with the VPF.
The VPF of a 4 deg$^2$ survey can distinguish between slow vs.\ fast, and early vs.\ late, reionization at $> 3-4\sigma$ at several redshifts between $7<z<9$. However, a survey of 13-16 deg$^2$ would allow the VPF to give several robust constraints ($>5-8\sigma$) across the epoch of reionization, and would yield a detailed history of the reionization of the IGM and its effect on Lyman-$\alpha$ Emitter clustering.
\end{abstract}


\section{Introduction} \label{sec:intro}

The epoch of reionization (EoR) is the era when the earliest galaxies ionized the ultraviolet-opaque `fog' of neutral hydrogen that filled the early universe.  Reionization history is still not well constrained, as various probes have led to different conclusions about its timing, pace, and sources \citep{McQuinn2016}. 
Lyman-$\alpha$ emitters (LAEs), which are star-forming galaxies that strongly emit in the Ly$\alpha$ line, offer practical probes of the reionization process.
% A probe for the EoR that will help clarify our understanding are Lyman-$\alpha$ emitters (LAEs), star-forming galaxies that strongly emit in the Ly$\alpha$ line. 
Detecting LAEs by their line emission enables surveys of otherwise faint galaxies across a wide redshift range.  Because the \lya\ line is resonantly scattered, it is easily attenuated by any amount of neutral hydrogen, and its visibility is highly sensitive to the ionization state of intergalactic gas throughout the reionization process (as predicted by \citealt{MiraldaEscude1998, HaimanSpaans1999}).
% Due to the line's resonance, especially at high-redshift and throughout the EoR, the emission of LAEs is very sensitive and easily attenuated by any amount of neutral hydrogen in the medium around them. 
%Therefore, the properties and distribution of LAEs can trace and describe the process of reionization (e.g.  RhoadsMalhotra2001, EMHu2002, Santos2004, MalhotraRhoads2004, Furlanetto2004, Furlanetto2006, McQuinn2007, Mesinger+Furlanetto2008, Sobral2015, Hu2021}). 

A common method of constraining reionization with LAEs uses the Ly$\alpha$ luminosity function (LF; first implemented by \citealt{MalhotraRhoads2004}; e.g.\ \citealt{Ouchi2010, Santos2016, Ouchi2018, Morales2021}).  This works by comparing the observed Ly$\alpha$ LF to that expected in a fully ionized medium (commonly based on the observed LF at a redshift where reionization is believed to be complete, e.g. $z=5.7$, \citealt{Ajiki2003, Ouchi2003, Hu2004}). 
LAE LF studies have reached various constraints for reionization. Some report suppression of the LAE LF in ranges within $5.7<z<7$, with inferred neutral fractions of order 20--40 percent (\citealt{ McQuinn2007, Kashikawa2006, Iye2006, Ouchi2010, Kashikawa2011, Konno2014, Konno2018}). Others report upper bounds on neutral IGM 
(\citealt{MalhotraRhoads2004, SnJ2006}), including some bounds that are tight enough to challenge some of the suggested detections of neutral gas whose results of $< 20$--30 percent at $z=7$ are consistent with a fully ionized medium]{Wold2022a}. 
%
%ranging from 
%a neutral fraction of 20--40 percent at $z\sim7$ (\citealt{Santos2004, McQuinn2007, Konno2018, Ouchi2010}), to an upper bound of 20--30 percent neutral \citet{Wold2022a}. Some LF studies have found no evidence for any neutral IGM attenuation at $z>6$ (\citealt{MalhotraRhoads2004, Tilvi2010}); while others measured an attenuation in the LFs or LAE number densities within $5.7<z<7$ (\citealt{Kashikawa2006, Iye2006, Ouchi2010, Kashikawa2011, Konno2014, Konno2018}).
%
LF suppression can have causes beyond a partially reionized IGM, such as cosmic variance, the evolution of the halo mass function \citep{Dijkstra2007}, and evolution in various factors that affect which galaxies strongly emit Ly$\alpha$ (e.g.\ \citealt{Ota2008, Stark2010, Pentericci2011, Ono2012, Schenker2012, Endsley2021, Hassan2021}). 

However, some constraints on reionization from the LAE LF conflict with those from other measurements. For example, results suggesting a mostly ionized universe at $z\sim 6$ (\citealt{Fan2006} using the IGM temperature and quasar \citealt{G-P1965} troughs), or $z>7-8$ extended reionization (from cosmic  microwave background anisotropies, \citealt{WMAP2009}). Studies examining the Ly$\alpha$ emission directly find mixed results, with some suggesting moderate to very high neutral fraction in the IGM at $z\sim 7.5$ (e.g. \citealt{Jung2020} v.s.\ \citealt{Hoag2019, Mason2019}), likely due to gaps in our understanding of how Ly$\alpha$ visibility during the EoR evolves in different types of galaxies and the inhomogeneity of reionization \citep{Jung2022}. Some results challengen even the the extremely early \citet{Finkelstein2019} model of reionization, such as the implied background of Ly$\alpha$ photons at $z\sim14$ to create the 78 MHz absorption profile of EDGES \citep{Bowman2018}, and the very recent detection of a high EW LAE at $z=10.6$ using JWST \citep{Bunker2023}.
There is still not a clear definitive picture of the neutral fraction of the IGM across these studies and methods; deep, blind searches for Ly$\alpha$ over large volumes offer a path to robust understanding.

The clustering of EoR LAEs is independent from the evolution of their intrinsic LF, and provides an additional method of constraining reionization to help resolve these tensions \citep{McQuinn2007}. Most constraints have used the angular two-point correlation function (\citealt{Ouchi2010, SobbachiMesinger2015, Ouchi2018, Gangolli2021}) or related statistics (e.g. count-in-cells in \citealt{Jensen2014}). In this work, we focus on the \textit{Void Probability Function} (VPF), a less common choice for EoR LAEs (\citealt{Kashikawa2006, McQuinn2007}) but one that may offer more sensitivity than the angular correlation function under some models \citep{Gangolli2021}. The VPF quantifies clustering by measuring how likely a circle or sphere of a given size is to be empty in a galaxy sample. As the 0$^{\text{th}}$ moment of count-in-cells, or zero-point volume-averaged correlation function, it carries the signature of higher order correlation functions. Its simplicity can be used to derive guidelines for the required density and volume of surveys \citep{Perez2021}. 

% In Perez et al.\ (2022a; henceforth Paper 1)
In \citet[][henceforth Paper 1]{Perez2022}, we quantify what constraints the VPF could yield for reionization when applied to
large-area blind narrowband searches for LAEs, such as the 
Lyman-Alpha Galaxies in the Epoch of Reionization (LAGER) narrowband survey. 
%The VPF has been studied for reionization with LAE from Subaru telescope observations at $z=6.6$ and $z=5.7$ (\citealt{Kashikawa2006,McQuinn2007,Gangolli2021}.
LAGER detects $z=6.9$ LAEs using a narrowband filter centered at 9642\AA\ with FWHM=92\AA\ (corresponding to approximately 30 cMpc at $z=6.9$), mounted on the 3.3 deg$^2$ field-of-view DECam on Cerro Tololo's 4-meter Blanco Telescope. So far, LAGER has yielded initial constraints of $\langle x_i \rangle_v > 0.67$ with its four-field LF (\citealt{Wold2022a}, building off \citealt{Hu2019}), and has the observation and analysis of another four fields in progress. Using the \citet{Jensen2014} simulations, Paper 1 
% predicts possible constraints for various iterations of LAGER with the VPF. Paper 1 
identifies how well 1, 4, or 8 LAGER DECam fields distinguish different ionization fractions with the VPF, and lays out a framework and case for using the two-dimensional VPF together to constrain the ionized hydrogen fraction implied by LAE clustering.

While narrowband surveys like LAGER (\citealt{Zheng2017,Hu2019,Hu2021,Wold2022a})
and SILVERRUSH (\citealt{Ouchi2018, Konno2018}) have been constraining the end of reionization with LAEs, the higher redshift EoR has been much more difficult to study on a large scale with LAEs. Ground-based infrared surveys are increasingly impractical at redshifts beyond $z=6.9$ (where reionization is thought to be mostly complete), both due to atmospheric OH emission creating prohibitively bright sky foreground at most wavelengths, 
% , which makes the sky foreground prohibitively bright at all but a few wavelengths, 
and a steep drop in silicon detector response at 1$\mu$m that restricts searches to instruments with comparatively small fields of view. Despite these challenges, focused ground-based searches for Ly$\alpha$ at $z=7.7$ and beyond (\citealt{Tilvi2020, Oesch2015, Zitrin2015}) have yielded detections of Ly$\alpha$ emitters.
%% JER comment We and others have searched at 7.7 and 8.8 from the ground. and $z=8.8$ (FLARE)

The Roman Space Telescope, NASA's next flagship mission set to launch in the mid-to-late 2020's, is an infrared telescope with a 2.4 meter Hubble-sized mirror, and a wide-field instrument with a 0.281 deg$^2$ field of view (200 times that of Hubble's WFC3-IR). Roman's wide-field instrument will have a slitless grism that is capable of capturing Ly$\alpha$ at $7.2<z<14$, as well as a lower-dispersion prism that will reach Ly$\alpha$ at lower redshifts ($z>6$). Roman can carry out surveys of reionization-era LAE clustering that will notably refine our understanding of the EoR, giving definitive constraints for how and when reionization occurred.

In this work, we adapt our framework from Paper 1 to make predictions for LAE clustering observations with Roman.  We quantify how Roman will constrain the timing and pace of reionization given its wide field, sensitivity, and continuous redshift coverage by combining three ingredients: a model for reionization in a $\sim (600 \hbox{cMpc})^3$ box that includes galaxy formation and radiative transfer of \lya\ \citep{Jensen2014}; a set of  reionization history models that we use to map between redshift and a given simulated neutral fraction; 
and instrumental sensitivity predictions based on detailed simulations of Roman grism data sets (Wold et al, in prep).   We then calculate the expected VPF and its expected uncertainties for surveys covering $\sim 1$, $\sim 4$, and $\sim 16$ square degrees, at each of five neutral fractions ($\langle x_i \rangle_v=$ \{0.22, 0.40, 0.485, 0.66, 0.93\}), in each of three reionization scenarios (gradual; rapid and early; rapid and late).  
We thereby address how large a clustering survey must be to distinguish between models for reionization history.

% To do this, we slice up the \citet{Jensen2014} (602$\times$607$\times$600) cMpc$^3$ simulations of LAEs 
%% in the volume-averaged ionized IGM fractions of $\langle x_i \rangle_v=$ \{0.22, 0.40, 0.485, 0.66, 0.93\} 
%at a range of neutral fractions ($0.22 \le \langle x_i \rangle_v \le 0.93$)
%to create mock Roman reionization-era LAE data sets. We consider narrow redshift slices with $\Delta z=0.2$, and identify the redshifts where various different models of reionization predict a given ionized hydrogen fraction. Next, we use the simulation results for Roman's grism sensitivity to Ly$\alpha$ for a deep 70 hour pointing (Wold et al., in prep), and create LAE samples for each redshift-reionization scenario. We measure the VPF across all samples and scenarios, and quantify 
% how constraints on reionization change with surveys of 1, 4, and 16 deg$^2$.
% what area is needed to distinguish models of reionization with Roman LAE clustering. 
% How large must a clustering survey be to identify the best description for our universe's reionization history--1 deg$^2$? 4 deg$^2$? Or up to 16 deg$^2$? 

This paper is laid out as follows. In \textsection \ref{sec:Sims_Models}, we describe the theory and simulations that support this work: the \citet{Jensen2014} simulations of LAEs through reionization in \textsection \ref{subsec:UsingJensenSims}; the models for reionization we compare in \textsection \ref{subsec:ModelDefs}; our application of the I.G.B. Wold et al.\ (2023, in preparation) Roman-Ly$\alpha$ grism simulations in \textsection \ref{subsec:IsakSims}; how we create mock Roman LAE surveys from the above in \textsection \ref{subsec:SimProjections}; and finally, how we apply the VPF for reionization constraints in \textsection \ref{subsec:MeasuringVPF}. In \textsection \ref{sec:VPFzevol}, we explore how the VPF evolves throughout the reionization history of the universe with different survey constructions, and how precisely Roman will be able to make out the epoch and pace of reionization.
In in \textsection \ref{subsec:z7p75focus}, we first focus on comparing the models at $z\sim7.75$ using the entire VPF as a function of clustering distance scale, VPF(R). We then
focus on an ambitious 13-16 deg$^2$ survey in \textsection \ref{subsec:FullFaceConstraints}, and smaller 1 and 4 deg$^2$ surveys in \textsection \ref{subsec:SmallerAreaConstraints}.
% ; and a focused $z\sim7.75$ model comparison for the entire VPF as a function of clustering distance scale VPF(R) in \textsection \ref{subsec:z7p75focus}. 
Appendix \ref{app:VPFdistributions} shows in detail our VPF measurements. We conclude and summarize our results in \textsection \ref{sec:Conclusion}.

%------------------------------------------------------------------------

\section{Simulations and Methods} \label{sec:Sims_Models}

% \textcolor{blue}{LUCIA THINKS: Maybe better order is: Roman sensitivity, models for reionization, Jensen simulations; slices to project Roman VPF measurements}

\subsection{Large Simulations of LAEs in Partially Reionized IGM} \label{subsec:UsingJensenSims}

For samples of LAEs at various ionization fractions, we use the \citet[J14]{Jensen2014} simulations of LAEs, which include radiative transfer through various amounts of ionized intergalactic medium (IGM). The simulations exist for discrete \textit{volume} ionized fractions of \{22, 40, 48.5, 66, 78, 89, and 93\} percent, corresponding to mass ionized hydrogen fractions of \{30, 50, 58, 73, 83, 92, and 95\} percent. 
% These simulations are chosen for their large volume, capable of accommodating some of the larger surveys that will be possible with Roman. 
We direct readers to Paper 1 for a more complete description of the J14 simulations and their properties for contexts similar to this work, but review relevant details for their assumptions for reionization (explained fully in \citealt{Jensen2013}). Finally, we note there has been exciting progress in improving the simulation of Ly$\alpha$ transmission during the EoR  (e.g.\ THESAN from \citealt{Thesan} and SPHINX from \citealt{SPHINX}, or \citealt{Qin2022} with the Meraxes semi-analytic model), though reaching volumes similar to the J14 simulations with these methods remains computationally difficult.

The underlying reionization model of the J14 simulations can be summarized as an inside-out reionization history that began somewhat early and progressed at a moderate pace. %This core scenario found the IGM moving from 40$\%$ ionized at $z=7.7$ to fully ionized at $z=6.5$ (Figure 2 of \citealt{Jensen2013}, whose reionization history resembles that of \citealt{Robertson2015}). 
They incorporated self-regulation as galaxies take part in reionization, turning off small sources of ionizing radiation once the IGM passed 10$\%$ ionization.
Briefly, halos were populated with galaxies as to match key luminosity functions in \citet{Ouchi2010}; each galaxy was modelled to intrinsically emit a halo mass-dependent double-peaked Ly$\alpha$ spectrum. Scattering from the IGM was incorporated using \text{IGMTransfer} based off higher resolution hydrodynamic simulations \citep{Laursen2011, IGMtransfer}.
% We use the discrete snapshots of different ionized fractions and apply redshift-dependent selections, for the redshifts prescribed by other reionization scenarios (slow/fast, early/late).
% The  LAE catalogs at the highest ionization fractions  ($\langle x_i \rangle_v > 0.6$)
% %, with no modifications or applied duty cycles) 
% are consistent with the luminosity function of $z=5.7$ LAEs \citep{Ouchi2010}.
% %JER comment: This seems wrong (lucia assumes JER is referring to the z-5.7 LF bit
% \textcolor{cyan}{NEW FROM LUCIA: }
J14 assumed that the luminosity function at $z=6.5$ from \citet{Ouchi2010} held true, and each of the `scenarios' correspond to the global-mass ionized fraction of the Universe at $z=6.5$. They constructed these `scenarios' by taking the single underlying reionization simulation at a given ionization fraction and scaling the halo masses until the intrinsic luminosity function matched that at $z=5.7$. 
% Our underlying assumption is that the intrinsic evolution of Ly$\alpha$ galaxies is small enough to be negligible during the period of reionization. For example, $\Delta z=0.5$ only corresponds to 65 Myr at $z=7$. This assumption may not hold as well when we try to reproduce slower reionization scenarios.
% The simulations model evolution of the halo mass function and galaxy luminosity function along with the IGM neutral fraction, so that differences in the galaxy population between different neutral fractions are automatically included in analyses with these catalogs. #---> lucia doesn't think this is actually true, hmm
We and J14 assume the intrinsic evolution of Ly$\alpha$ galaxies is modest during the rapid phases of reionization (for example, $\Delta z=0.5$ only corresponds to 65 Myr at $z=7$). 

% \textcolor{blue}{\textbf{Lucia wonders:} should we include the LFs of the simulations? To be very clear about our assumptions?}

Later in this work, we redshift the LAE catalogs to reflect the redshifts  different models for the EoR predict for each ionized IGM fraction (i.e.\ slow/fast, early/late reionization), and then consider what Roman will observe in the context of the VPF of LAEs. We primarily focus on the $\langle x_i \rangle_v$=\{22, 40, 48.5, 66, 93\} percent ionized simulations. Most of these fractions are predicted to exist at redshifts where the Roman grism will observe Ly$\alpha$ under the models of reionization that we analyze. The most ionized volume is in particular a useful baseline of clustering, and can be compared to ground-based surveys as well as future studies with the Roman prism that will detect Ly$\alpha$ to $z>6$. Finally, we note that when we redshift the luminosities, we are maintaining the core result of the J14 simulations–-how LAEs in different fractions of ionized IGM under their model appear, and how different ionization fractions affect the large scale structure as independently of LF evolution as possible--and then incorporating the straightforward dimming due to redshift.

Finally, we note that our application of the J14 simulations here differs in one notable way from our work in Paper 1. Paper 1 specifically focused on creating predictions for the VPF of the observed LAGER $z=6.9$ LAEs, and our selection reflected this by applying a realistic LAGER-like luminosity-dependent selection and a LAE duty cycle (following \citealt{Kovac2007}) to yield good consistency with the observed LAGER LAE density and luminosity function. However, as we broaden our analysis to Roman observations across such a broad redshift range, where the underlying UV and \lya\ luminosity functions have not yet been constrained with observations, we do \underline{not} apply a duty cycle, and use the J14 LAE catalogs with no modifications. 
Once these luminosity functions are better understood (e.g.\ with JWST), a retroactive duty cycle can be applied to our results by scaling all predicted number densities and all log$_{10}$(VPF) measurements by the duty cycle.

\begin{figure}
	\begin{center}
\includegraphics[width=0.85\textwidth]{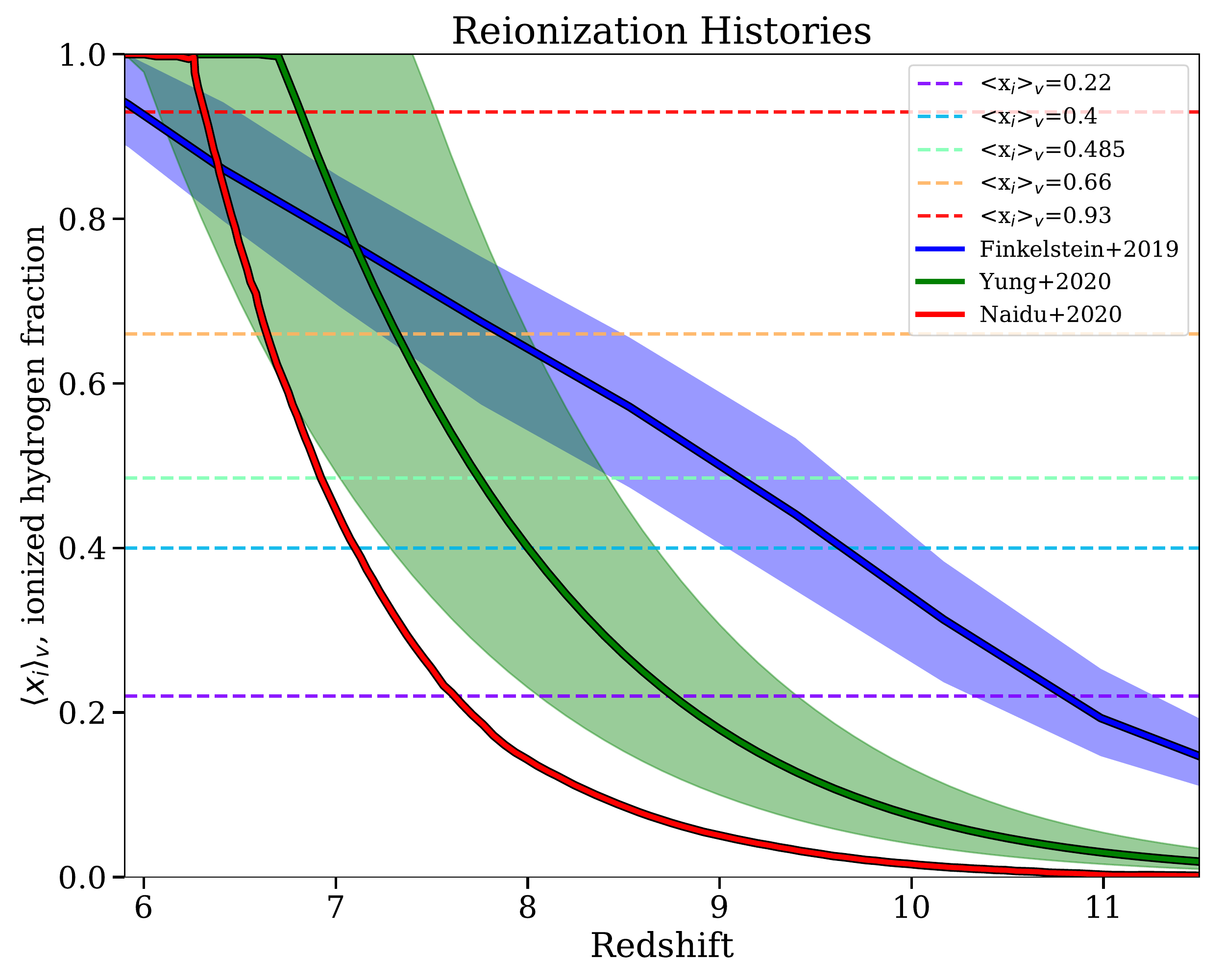}
	\caption{The predicted volume-averaged ionized hydrogen fraction by redshift according to several models of reionization: \citet[F19, blue]{Finkelstein2019}, \citet[Y20, green]{Yung2020}, and \citet[N20, red]{Naidu2020}. We identify at which redshift the models (solid lines) predict the ionization fractions simulated by J14 (dotted horizontal lines).
	\label{fig:ReionizModels}}
	\end{center}
\end{figure}

\subsection{Models for Reionization} \label{subsec:ModelDefs}

We focus on a representative sample of models for the reionization history of the universe to guide our projections for Roman. Figure \ref{fig:ReionizModels} shows the redshift vs.\ volume-averaged fraction of ionized hydrogen in the IGM, according to the reionization models of \citet[F19]{Finkelstein2019}, \citet[N20]{Naidu2020}, and \citet[Y20]{Yung2020} (respectively, blue, red, and green). Figure \ref{fig:ReionizModels} shows the discrete ionization fractions of the J14 simulations as dashed horizontal lines ranging from most neutral (purple) to most ionized (red). 

These models sample a range of reionization histories, driven by different assumptions about the production of ionizing photons.
% most have not yet been confidently ruled in or out with current constraints. 
These models are tuned to reproduce existing constraints for reionization, such as:
% all of these models reproduce the electron scattering optical depth measured with the CMB, and IGM ionization constraints from $z\sim7$ quasars. 
measurements of the IGM temperature that indicate a mostly ionized IGM at $z\sim6$ (e.g. \citealt{Fan2006}); analyses of measured Ly$\alpha$ line profiles that rule out a fully neutral universe at $z=6.6$ (\citealt{HaimanCen2005, Ouchi2010, Rhoads2013}); quasar \citet{G-P1965} trough studies that find a fully ionized universe by $z=5-6$ (e.g.\ \citealt{Fan2006, McGreer2015}); and measurements of the average Thomson scattering optical depth with cosmic microwave background (CMB) anisotropies (e.g.\ \citealt{WMAP2009, Planck2016}) that support higher-redshift extended reionization ($z>7-8)$. These models also focus on galaxies as the likeliest and dominant sources of reionization (e.g.\ \citealt{Robertson2013}), supported by the rarity of quasars beyond $z>6$ and constraints on helium reionization (e.g.\ \citealt{MadauHaardt2015}). Next, we briefly describe these models for reionization.
    
F19 present a semi-empirical model of reionization whose core is a physically-motivated halo mass-dependent parametrization of escape fractions. In their model, the faintest galaxies in the UV collectively dominate the ionizing emissivity, leading to a reionization history that starts very early (80$\%$ volume-ionized at $z\sim7$) and progresses at a smooth pace. They also model an AGN contribution to the end of reionization, making up one third of the budget at $z=6$, and predict a flat star formation rate density at $z>8$.

Y20 use end-to-end semi-analytic models with the goal of modeling all ionizing sources in fine detail. They use physically motivated relationships between dark matter halo formation histories and galaxy properties (including synthetic spectral energy distributions) to connect galaxy formation physics to the large-scale reionization history. They combine the Santa Cruz semi-analytic model for galaxy formation (\citealt{SomervillePrimack1999, Somerville2008, Somerville2015}) with an analytic reionization model (\citealt{Madau1999}, similar to that in \citealt{Naidu2020}), and have only the escape fraction as a free parameter. 
% Their analytic model for the reionization of hydrogen in IGM is from \citet{Madau1999}, and is quite similar to that of N20. 

N20 create and apply an empirical model to explore what objects carried out the bulk of ionization. Focusing specifically on ionizing photon escape fractions, their model ties the escape fraction to the star formation surface density (as motivated by recent samples of Lyman continuum leakers). Their model implies that rare very massive and UV bright galaxies (`oligarchs') account for the vast majority of the reionization budget. 

When comparing these models purely on the reionization history they predict in Figure \ref{fig:ReionizModels}, N20 and Y20 follow a similar pattern: reionization starts very slowly and ramps up quickly after $z<8$. N20 stands out among many reionization models for how late and rapidly it occurs. F19 also stands out by presenting a reionization history that began earlier than $z>11$ and evolved slowly and steadily. In the context of LAE observations with Roman, a universe that F19 describes will find many more LAEs at higher redshifts, as the IGM has more ionized gaps that allow the photons through. On the opposite side, N20 would predict very few LAEs at $z>7$ and a mostly neutral IGM. Roman will hugely inform our understanding of reionization, with its large field-of-view and vast redshift coverage beyond the low-redshift universe where ground based observations have not found definitive constraints between these models.

% Finally, though we do limit ourselves to these three models for analysis, we also note a few other models with differing reionization histories and explanations for reionization. \citet{Kulkarni2019} and \citet{Robertson2015} both have a similar `ramp-up' shape as do \citet{Yung2020} and \citet{Naidu2020}, but  \citet{Robertson2015} starts the earliest of them while \citet{Kulkarni2019} ends the latest.
% \citet{Kulkarni2019} ran a radiative transfer simulation of cosmic reionization by galaxies, with the express goal of reproducing the observed Ly$\alpha$ forest opacity. They found a very late reionization--reaching complete ionization after $z<6$-- due to the emissivity model they use in order to match the Ly$\alpha$ opacity distribution.%: their model has peak emissivity at $z=6.8$, and find it decreases at higher redshift more slowly than predicted by the evolution of the cosmic star formation rate density. 
% \citet{Robertson2015} instead assessed the assumption that most of the photons that reionized the universe came from high redshift star forming galaxies. Their empirical inputs used simple assumptions, like a constant escape fraction. This models predicts very quick reionization between $6<z<9$, does not require many $z\gg10$ galaxies, and supports the assumption that star-forming galaxies dominated reionization.

\subsection{Expected Roman Grism Ly$\alpha$ Response} \label{subsec:IsakSims}

\begin{figure}
	\begin{center}
    \includegraphics[width=0.75\textwidth]{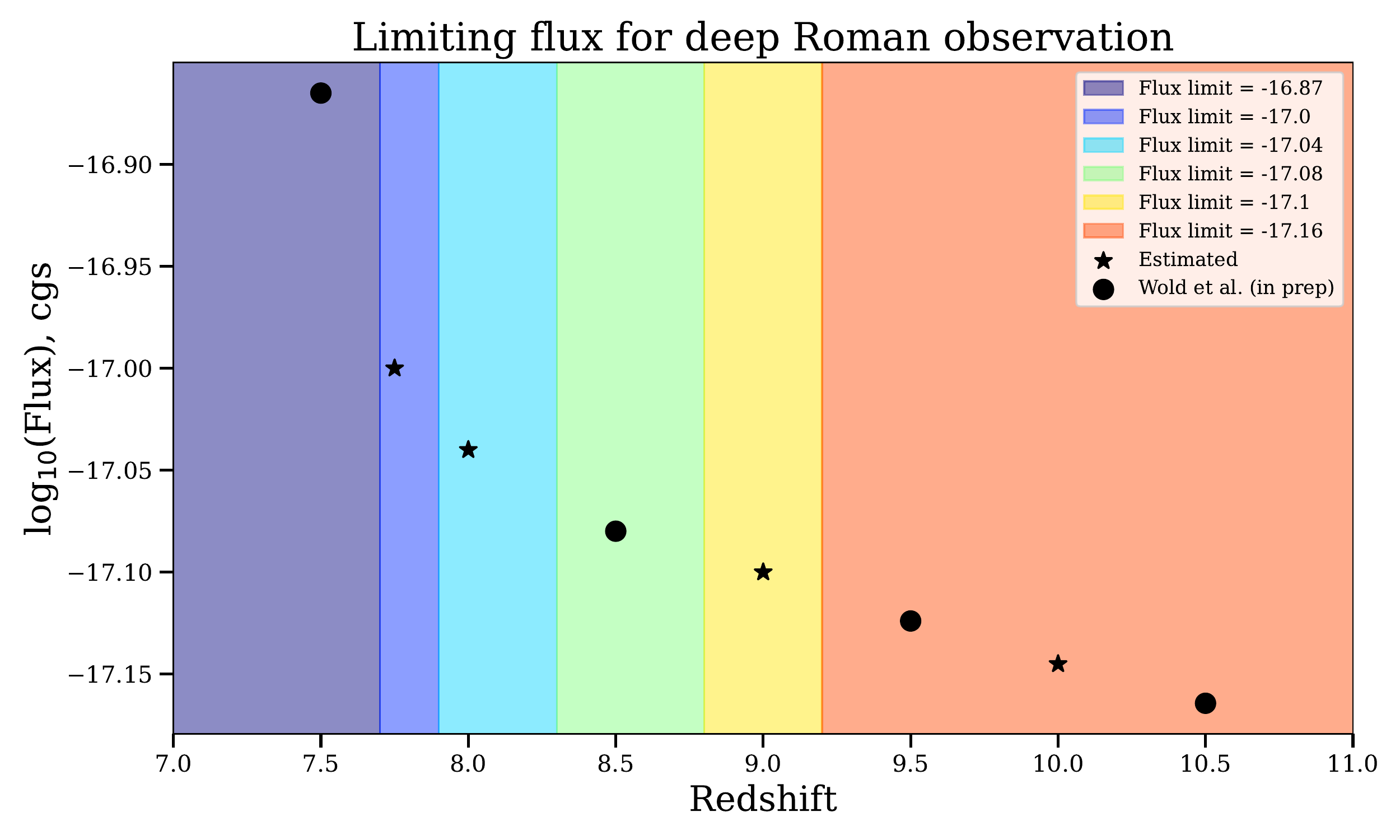}
	\caption{The redshift-dependent flux limits we use, as informed by the detailed Roman Ly$\alpha$ simulations of Wold et al.\ (2023, in preparation). The thresholds for applying each flux are indicated by the blocks of shaded color (shallowest at the lower redshifts in indigo, to the deepest high redshift observations in orange).
	Circles show the fluxes where 50$\%$ completeness of LAEs was directly simulated for a deep imaging; stars are our interpolations for redshifts in between. We treat the completeness behavior as a step-function. The fluxes applied to each redshift-reionization scenario are listed in Table \ref{table:AllModels}.
	\label{fig:zdepfluxes}}
	\end{center}
\end{figure}

\begin{figure}
	\begin{center}
    \includegraphics[width=0.7\textwidth]{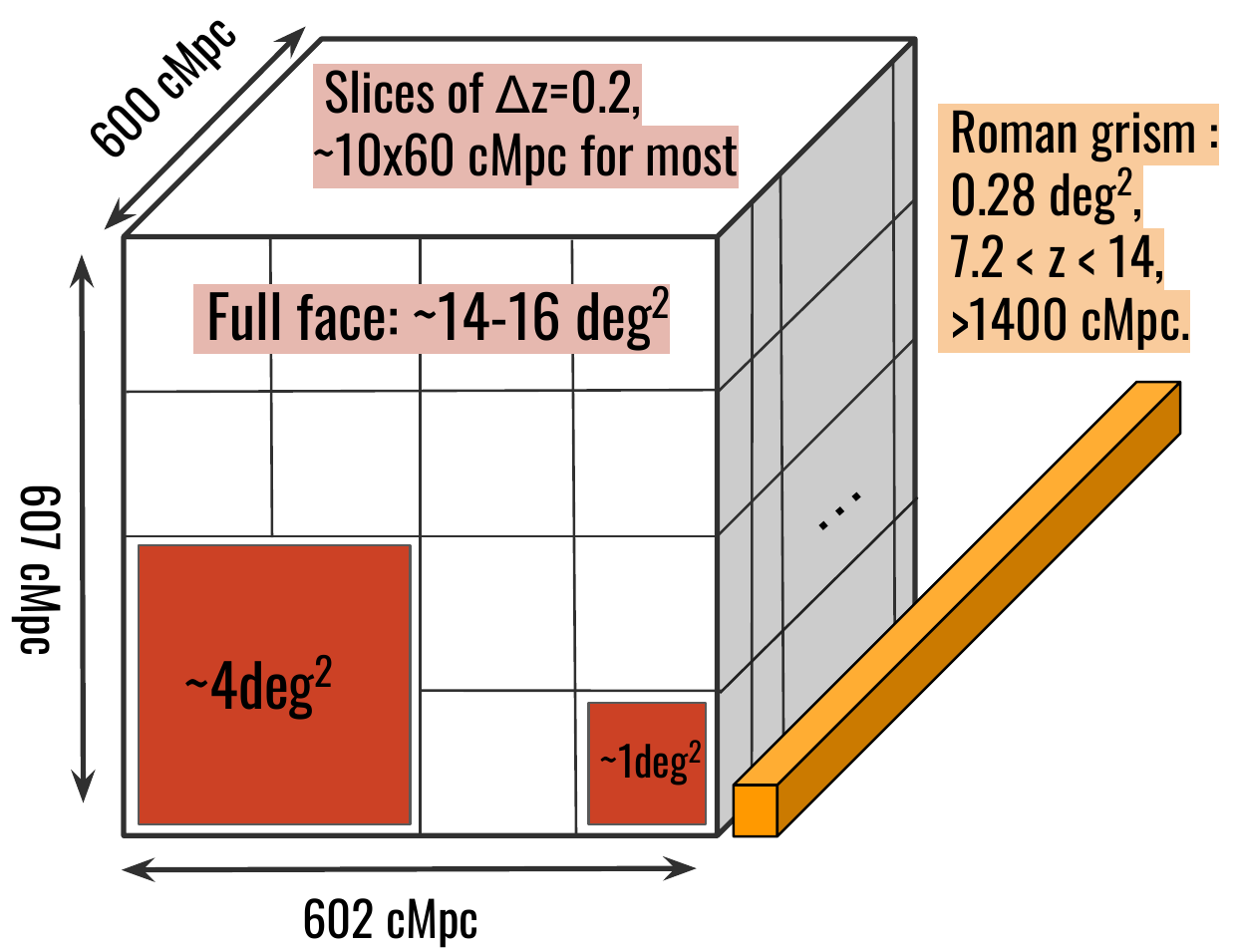}
	\caption{A schematic showing how we slice the J14 simulations to explore the constraints on reionization the Roman Space Telescope may find with the VPF. Depending on the redshift studied, the full face of the (601x607x600) cMpc$^3$ volume corresponds to approximately 13-16 deg$^2$. For our exploration of how the VPF evolves by redshift under different models for the reionization history of the universe, we test three survey areas with Roman: the full face $\sim16$ deg$^2$, quarter-face $\sim4$ deg$^2$, and sixteenth-face $\sim1$ deg$^2$. We assume a survey covering $\Delta z$=0.2, which yields between 7 to 14 equally deep slices in the simulations. For comparison, we show in yellow a long thin tube that illustrates the area and Ly$\alpha$ redshift coverage for one imaging (or, a single FOV pointing) with the Roman grism. Table \ref{table:AllModels} detail each redshift-reionization scenario's slice depth and approximate LAE surface density. \label{fig:CutUpSims}}
	\end{center}
\end{figure}

% \begin{table*}[t]
% 	\begin{center}
%     \caption{
%     The redshift-dependent flux limits we use, as informed by the detailed Roman Ly$\alpha$ simulations of Wold et al. (2022b) and the redshifts where the \textsection \ref{subsec:ModelDefs} reionization models predict the \citet{Jensen2014} ionization fractions. \label{table:zdepfluxes}
%     }
% 	\begin{tabular}{p{2.5cm}p{3.5cm}} 
% 	\hline \hline
%     Redshift range & Limiting Flux, erg sec$^{-1}$  \\
%     \hline
% 		$z<7.7$      & 1.4$\times10^{-17}$ \\
%         $7.7<z<7.9$  & 1.0$\times10^{-17}$ \\ %$7.7<z<8$   & 1.0$\times10^{-17}$ \\
%         $7.9<z<8.3$  & 9.1$\times10^{-18}$ \\ %$8<z<8.8$   & 8.3$\times10^{-18}$ \\
%         $8.3<z<8.8$  & 8.3$\times10^{-18}$ \\
%         $8.8<z<9.2$  & 7.9$\times10^{-18}$ \\
%         $z>9.2$      & 6.9$\times10^{-18}$ \\
%     \hline \hline	
% 	\end{tabular}
% 	\end{center}
% \end{table*}
% %% JER comment: Lose some digits in table 1.

The final step needed for our work is Roman's expected sensitivity to Ly$\alpha$ across its broad redshift range.
% The next consideration to creating projections of Roman reionization observations is approximately how deeply it will observe Ly$\alpha$ across its broad redshift range.
% To make the most helpful predictions for Roman observations of LAEs, 
Here, we leverage some of the core results of Wold et al.\ (2023, in preparation).
% for the Ly$\alpha$ detection limits Roman will reach. 
That work carries out detailed simulations of LAE identification in Roman grism data, and quantifies the expected completeness of LAE samples across a range of luminosity and  redshift. For this work, we use the results of their deepest simulation: 25 position angles at 10 kiloseconds each, %  for 14.1 arcmin$^2$ field-of-view,
% \footnote{The analysis of Wold et al.\ (2023, in preparation) is limited to this area, roughly one quarter of a Roman chip, due to available HST observations of the COSMOS field.}
for a total of 70 hours of exposure. Wold et al.\ (2023, in preparation) additionally assume that a deep grism field would cover the same area as a deep broad-band field, meaning low-redshift emission line galaxies would be easily identified and lead to a very low contamination rate in the final LAE sample.
% the Wold et al FoV is not relevant here.

Wold et al.\ (2023, in preparation) test four bins for redshift completeness
% over $\Delta z=0.25$ centered 
at $z$=7.5, 8.5, 9.5, and 10.5 for the Roman grism. They insert LAEs with known line flux and measured their recovered fraction.
% assume the $z=6.9$ LAGER Ly$\alpha$ luminosity function is true at these redshifts, with no galaxy evolution or reionization \citep{Wold2022a}. 
% They populate the field with false LAEs at the central redshift, and record how many are detected above a target flux cutoff in the presence of a detailed and realistic extra-galactic foreground scene. 
% With this process, t
They have quantified what flux and Ly$\alpha$ line luminosity limits correspond to 50$\%$ completeness for various key redshifts. For $z=\{ 7.5, 8.5, 9.5, 10.5\}$, the limiting fluxes are \{1.4e-17, 8.3e-18, 7.5e-18, 6.9e-18\} erg s$^{-1}$ cm$^{-2}$, corresponding to Ly$\alpha$ line luminosities of \{9.1e+42, 7.4e+42, 8.6e+42, 9.9e+42\} erg s$^{-1}$ respectively. The completeness functions approximate step functions, and we treat them as so when applying these flux selections to the unmodified J14 catalogs.

% We use these results of Wold et al (2022b) as a redshift-dependent flux limit for a deep Roman survey. 
We show the redshift-dependent flux limits that we derive from Wold et al.\ (2023, in preparation) in Figure~\ref{fig:zdepfluxes}.
% We note that the transition from 0 to 100$\%$ complete is quite sharp in their results, so we treat the 50$\%$ cutoff as a step function. 
According to these simulations, Roman reaches its deepest limiting fluxes at the highest redshifts for Ly$\alpha$. The grism is least sensitive at its blue edge (near 1 $\mu$m, or $z=7.2$ for \lya) but becomes much more sensitive toward redder wavelengths, reaching as deep as $\log_{10} f_{Ly\alpha} >$ -17.15 cgs for $z=10.5$. However, when also considering the increasing luminosity distance of a given LAE, we expect the Roman grism to be most sensitive to \lya\ near $z=8.5$.   % , \textit{though future work can be updated to reflect simulations for the Roman prism sensitivity once they exist}.
Though the Roman grism will not observe any \lya\ below $z<7.2$, the Roman prism will observe \lya\  to $z>5.2$. Based on the results of Wold et al.\ (2023, in preparation) and this expected synergy between the Roman instruments, we choose to apply a ceiling sensitivity of 1.4e-17 erg s$^{-1}$ cm$^{-2}$ for all redshifts under $z<7.7$. Our assumption of this fixed flux limit at $z<7.7$ is conservative, since the prism has higher throughput than the grism at $\sim 1\mu$m. 

\subsection{Creating Mock LAE Samples for Various Roman Surveys} \label{subsec:SimProjections}

We now move to predicting the LAE clustering analyses that Roman surveys will enable, using the J14 simulated LAEs. 
% In \textsection 2.1 and 2.2, we described which of the J14 simulations will likely be observed by Roman, and when the models of reionization predict each discrete ionization fraction. In \textsection 2.3, we discussed the expected sensitivity of Roman to Ly$\alpha$ across its redshift range (Wold et al., in prep). 
% To probe how the VPF could constrain between these ionization models, we first 
% extract slices from the J14 simulations corresponding to the LAE clustering analyses that Roman surveys will enable. In particular,
In this section, we consider narrow slices of the J14 simulations (for a 2D VPF clustering analysis) and three different Roman survey constructions: $\sim1$ deg$^2$, $\sim4$ deg$^2$, or $\sim16$ deg$^2$.

Figure \ref{fig:CutUpSims} illustrates how we slice up the J14 volumes. We assume focused studies of depth $\Delta z \approx0.2$,
% centered at the redshift of focus,
% , following the \textcolor{purple}{\textit{recommendation of the RST Science Team(?)}}. 
and create equal slices across the 600 cMpc volume depth. As we first implemented in Paper 1, we prioritize creating fully independent slices from the volume, for a clearer sense of the behavior of the VPF for a given area. Depending on the redshift that a given ionization model predicts an ionization fraction, there can be between 14 slices $\times$ 43 cMpc/slice (e.g. F19 predicting $\langle x_i \rangle _v=0.22$ at $z=10.8$) to 7 slices $\times$ 86 cMpc/slice (e.g. F19 predicting $\langle x_i \rangle_v=0.93$ at $z=5.95$) in the $z$-direction. The columns under `Slice$\times$Depth' in Table \ref{table:AllModels} detail this for each redshift-ionization scenario. 
% Finally, we note that the Roman grism will observe Ly$\alpha$ mostly between $7.2<z<14.9$, yet we list lower redshift and higher ionization fraction details later in this work for observers analyzing LAEs at lower redshifts with the Roman prism and other observatories.

Now armed with mock $\Delta z\approx 0.2$ slices of the J14 simulations for each reionization model and redshift scenario, we apply a redshift-dependent flux limit (Figure \ref{fig:zdepfluxes}, \textsection \ref{subsec:IsakSims}) to the trasmitted \lya\ luminosity of each LAE in each slice.
% We note that our limits are not the exact interpolations of the results of the Wold et al. (2022b) simulations, but simplified guidelines based on those results as well as the redshifts where the models predict the ionization fractions.
%%JER: what does the preceding sentence mean?
% The simulated LAEs of J14 come with transmitted Ly$\alpha$ luminosities.  
In order to apply the Roman-specific flux limit from Figure \ref{fig:zdepfluxes} to a given LAE, we translate the given flux limit to a luminosity cut based on a given LAE's position within the slice. In each J14 full-face slice, we imagine the center $Z$-position is at the redshift of focus, with $\Delta z \pm 0.1$ as the back and front edges of the slice. We associate the relative comoving position of a LAE in the slice to a luminosity cut, and apply the given flux limit for the central redshift in Figure \ref{fig:zdepfluxes} to generate a flux-limited samples for 2D clustering.
% After all LAEs in a slice have had this flux-derived luminosity cut applied, we apply a duty cycle, randomly sampling 12.5$\%$ of the LAEs\footnote{LAEs are thought to have duty cycles, or a timescale determining what subset of halos host actively emitting LAEs, of likely 10-20$\%$ (e.g.\ \citealt{MalhotraRhoads2002, Kovac2007, Nagamine2010, Zheng2010, SobbachiMesinger2015, Hong2019}). This particular value finds good consistency with $z=6.9$ LAE observations \citep{Wold2022a}.}. 

% Or: we slice the volume into even slices; assuming the Z-center of the slice is the redshift of focus, the front end is zcenter -0.1, back is zcenter+0.1, tie comoving transverse distance to a redshift within range; get luminosity distance for all tiny steps in redshift; apply the flux limit from Table \ref{table:zdepfluxes} for the central redshift; so now the relative position in the slice is associated with a luminosity cut from the flux limit. For each full-face slice, go through each LAE: does it pass the flux cut corresponding to its relative position/redshift? Do for all slices; add up LAEs that passed all flux cuts, apply DC in each, and this gets 'total flux-limited LAEs' in Tables 3-6. For 4 or 1deg2, cut up post flux-cut and DC full face slices.
%% JER: the preceding sentence here is important, may warrant clarification.

Next, we consider the $X$ and $Y$ dimensions of our mock LAE samples. At the redshifts examined, the full-face of the (602$\times$607) cMpc$^2$ J14 simulations cover between 13-16 deg$^2$. We split each slice into exact fourths (301$\times$303.5 cMpc$^2$) or sixteenths (150.5$\times$151.75 cMpc$^2$) to examine slices of approximately 4 deg$^2$ or 1 deg$^2$, respectively. This maximizes the number of completely independent mock LAE samples we are able to create, and covers a broad range of possible survey areas. Later in this work, we use all independent $\Delta z=0.2$ slices to explore the variability in the VPF as a function of survey area.
% We use these slices to explore how constraints on reionization with the VPF and Roman change with survey area, and to guide future observations and analyses. 
Figure \ref{fig:CutUpSims} displays our handling of the J14 simulations, with the additional illustration of what the default Roman grism will observe: an area of 0.281 deg$^2$ simultaneously covering $7.2<z<14$. Therefore, Roman will need approximately 4 independent and non-overlpaping FOV pointings to cover 1 deg$^2$, approximately 16 for 4 deg$^2$, etc. Finally, though we are artificially creating slices of $\Delta z=0.2$ about specific redshifts, in truth Roman will be able to access LAEs across the bulk of reionization history. 

Table \ref{table:AllModels} summarizes the results of transforming each model into a redshift-reionization scenario for clustering measurements, including the size of the given slices, the limiting line flux applied, and resulting LAE surface density. To measure the surface density ``$\Sigma$, LAEs deg$^{-2}$", we average the number of the LAEs that pass their redshift-position flux cut across all full-face slices, and divide by the area implied for 602$\times$607 cMpc$^2$ at the central redshift. These are the LAE surface densities expected under the Wold et al.\ (2023, in preparation) sensitivity predictions and nuances of the J14 simulations\footnote{Assuming that the UV and Ly$\alpha$ LFs have \textit{not} evolved since $z\sim6$. As discussed in \textsection \ref{subsec:UsingJensenSims}, a duty cycle can be introduced into our results once the UV and/or \lya\ luminosity functions are better understood at these redshifts.}
across all our redshift-reionization scenarios, ranging from a few dozen to a few hundred deg$^{-2}$.

%----------------------------------------------------------------------------

\begin{table*}[t]
	\begin{center}
    \caption{Details for mock Roman observations given the \citet[F19]{Finkelstein2019}, \citet[Y20]{Yung2020}, and \citet[N20]{Naidu2020} models for reionization, generated from the unmodified \citet[J14]{Jensen2014} LAE catalogs. For a given $\langle x_i\rangle _v$ simulation, we list: the redshift each model predicts the ionized fraction; the number and depth of slices corresponding to $\Delta z=0.2$ depth; the flux limit applied at the redshift in log$_{10}$ erg sec$^{-1}$; the corresponding limiting Ly$\alpha$ line luminosity in a slice; and the average surface density in LAEs per deg$^{2}$ for a full-face slice.}
	\begin{tabular}{p{.9cm}p{.65cm}p{.65cm}p{0.65cm}p{0.75cm}p{0.85cm}p{.75cm}p{0.75cm}p{0.85cm}p{0.7cm}p{0.7cm}p{0.75cm}p{0.8cm}p{0.7cm}p{0.7cm}p{0.7cm}} 
	\hline \hline
	\multicolumn{1}{c}{$\langle x_i\rangle _v$} & \multicolumn{3}{c}{Slice Center $z$} & \multicolumn{3}{c}{Slices$\times$Depth, cMpc} & \multicolumn{3}{c}{Lim. Flux, log$_{10}$ erg s$^{-1}$} & \multicolumn{3}{c}{Lim. log$_{10}$L$_{\text{Ly}\alpha}$} & \multicolumn{3}{c}{$\Sigma$, LAEs deg$^{-2}$} \\
    \hline
    Model: & F19 & Y20 & N20 & F19 & Y20 & N20 & F19 & Y20 & N20 & F19 & Y20 & N20 & F19 & Y20 & N20 \\
    \hline
		0.22  & 10.8 & 8.75 & 7.6  & 14x43 & 12x50 & 10x60 & -17.2 & -17.08 & -16.9 &  43.04 & 42.91 & 42.98 &   29 &  52 & 43 \\
        0.40  & 9.65 & 8.0  & 7.1  & 13x45 & 10x60 &  9x67 & -17.2 & -17.04 & -16.9 &  42.92 & 42.88 & 42.91 &   66 & 103 & 85 \\
        0.485 & 9.1  & 7.75 & 6.9  & 12x50 & 10x60 &  8x75 & -17.1 & -17.0  & -16.9 &  42.93 & 42.89 & 42.88 &  104 & 148 & 160 \\
        0.66  & 7.85 & 7.3  & 6.65 & 10x60 &  9x67 &  8x75 & -17.0 & -16.9  & -16.9 &  42.88 & 42.97 & 42.85 &  162 & 133 & 211  \\
        0.93  & 5.95 & 6.85 & 6.3  &  7x86 &  8x75 &  8x75 & -16.9 & -16.9  & -16.9 &  42.73 & 42.9  & 42.79 &  381 & 200 & 273 \\
    \hline \hline	
    \label{table:AllModels}
    \end{tabular}
	\end{center}
\end{table*}
 
\subsection{Using the Void Probability Function for Reionization} \label{subsec:MeasuringVPF}

% \textit{\textcolor{red}{Is this the best place for this section? Perhaps it should start off the next one? Worry it could distract from the message in that section, though}}

The Void Probability Function (VPF) is a measurement of clustering that quantifies how likely a region of a given size is to be empty within the sample. Or, if one drops a certain number of circles of some $R$, how many are empty? It is the zero-eth moment of count-in-cells, focusing only on the cells with no galaxies, and therefore carries the signature of higher order volume-averaged clustering statistics (\citealt{White1979}). \citet{Perez2021} applied an analysis of the hierarchical scaling between the VPF and higher order correlation functions (see also \citealt{Conroy2005}),  comparing the VPF-derived volume averaged two-point correlation function and standard \citet{L-S1993} correlation function for simulated $3.1<z<6.6$ LAEs. They also derived theoretical descriptions for minimum and maximum distance scale guidelines for the VPF that we use in this work. 

The VPF has also been leveraged specifically for reionization constraints. \citet{Kashikawa2006} and \citet{McQuinn2007} examined the ability of the VPF to constrain reionization for Subaru LAE observations at $z=6.6$. More recently, \citet{Gangolli2021} explored the constraining power of the VPF and other clustering statistics for SILVERRUSH LAEs at $z=5.7$. In Paper 1, we used the VPF to quantify what constraints on reionization the LAGER narrowband survey will be able to make at $z=6.9$ with one, four, or eight fields' worth of imaging. 

We measure the VPF using the $k$-nearest neighbors algorithm introduced in \citet{Banerjee2021}, which is notably faster than other methods. In a given slice, the transverse comoving positions of LAEs that passed the flux cut are normalized to cover e.g. $x=0-301$, $y=0-303.5$, and $z=0-60$ cMpc for a $\sim4$ deg$^2$ slice at $z=9$. For each of several radii between 5 and 50 cMpc, we drop 100,000 random points in a slice to measure the VPF. We repeat this process 5 times to minimize sampling error, and use the average across the five samplings as the measured VPF of a given slice. When we later explore the VPF behavior of a given redshift-reionization scenario as a function of area, we will compare the mean and standard deviation of the VPF across all $\Delta z=0.2$ slices of a particular area.

For a focused statistical analysis, in this work we will primarily discuss the VPF at $R\sim12$ cMpc. $R\sim12$ cMpc roughly corresponds to expected scale of ionized bubbles during reionization for the redshifts we examine. The smallest visible bubbles in the Ly$\alpha$ distribution are ones that will results in a \citet{G-P1965} line center optical depth of $\tau \approx 1$, and which are therefore large enough to allow sufficient transmission of Ly$\alpha$ (see \citealt{RhoadsMalhotra2001, Rhoads2007}). The characteristic radius of these bubbles comes out to 1.2 physical megaparsec, or:
% The inside of a bubble is completely ionized and the outside is completely neutral, so that the line center optical of Ly$\alpha$ due to the damping wing from the IGM is $\tau=1$,

\begin{equation}
    R_{\text{bubble}},\ \text{cMpc}=1.2\ \text{pMpc} \times (1+z).
\end{equation}

The VPF is a volume-averaged clustering statistic, which in practice means that general trends are consistent between nearby distance scales. 
% \textcolor{cyan}{To first lay stage, let us first examine the VPF of the reionization models across all radii for the models near $z=7.75$.}
In \textsection \ref{subsec:z7p75focus}, however, we study the VPF across all radii for the models near $z\sim7.8$. %The VPF is inherently connected to higher-order correlation functions \citep{White1979, Maurogordato1987, Perez2021}, and its simplicity to implement and slight constraining edge over the angular correlation function \citep{Gangolli2021} make it a worthwhile tool for clustering studies of reionization-era galaxies.

%------------------------------------------------------------------------------

\section{Redshift Evolution of the VPF Across Reionization \label{sec:VPFzevol}}

\subsection{Example Focused Constraints near $7.6<z<7.9$} \label{subsec:z7p75focus}

\begin{figure}
    \centering
    \includegraphics[width=\textwidth]{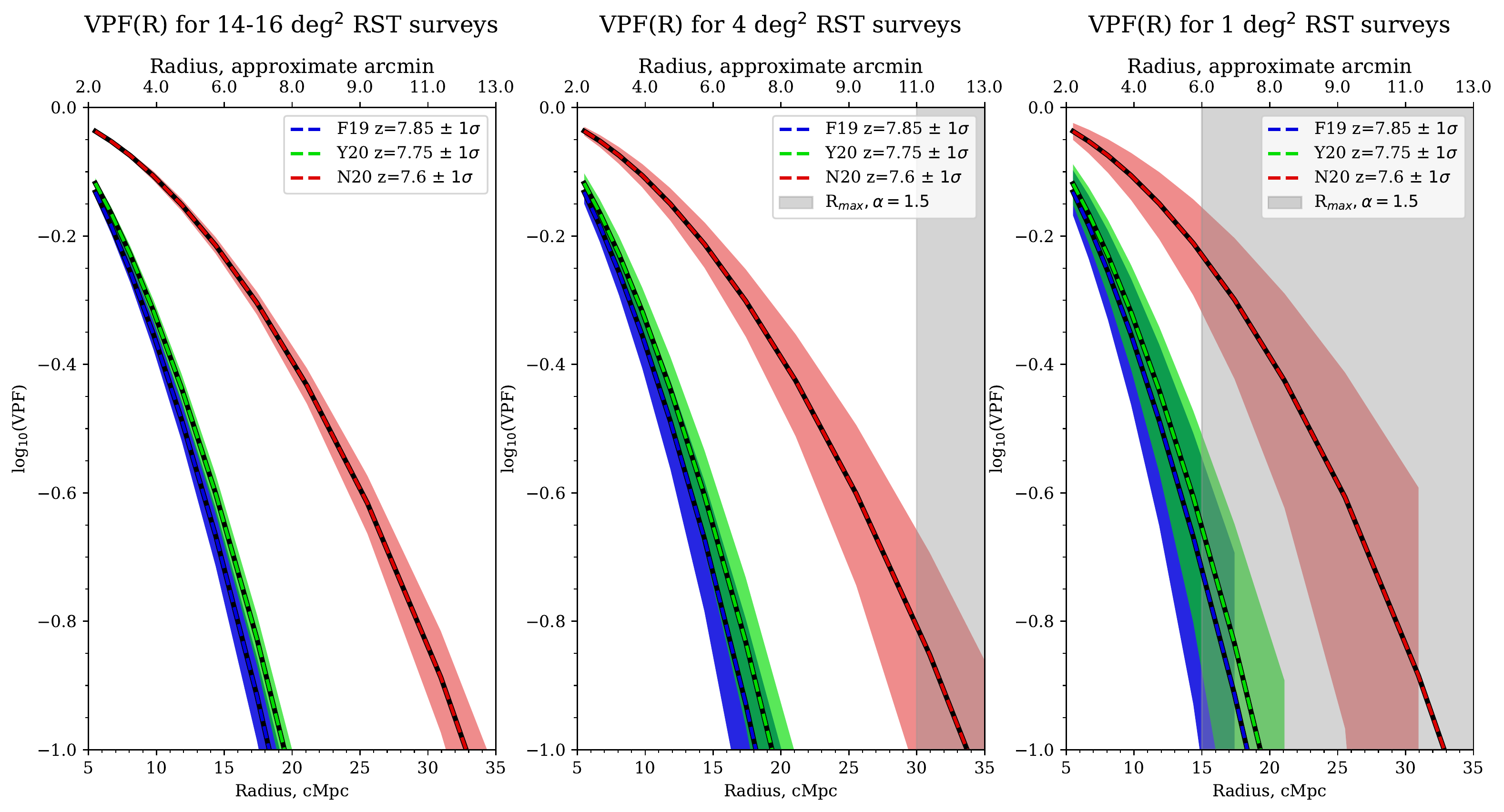}
    \caption{The complete VPF curves for the three ionization fractions--$\langle x_i \rangle _v$=0.22, 0.485, 0.66--predicted near $z\sim7.75$ by the three models (N20, A20, F19 in red, green, blue respectively). The colored sharing are the 1$\sigma$ standard deviation across all independent $\Delta z=0.2$ slices of the given area. Grey shading indicates the approximate distance scales where the given survey area cannot measure the VPF (assuming $\alpha=1.5$, and that all models exist at $z=7.75$ for simplicity). We note that for the full-face 14-16 deg$^2$ surveys, the blue F19 and green Y20 1$\sigma$ shadings begin to overlap only at $R>15$ cMpc; and, that for the 4 deg$^2$ surveys, the mean Y20 VPF values are similar to the upper 1$\sigma$ bound of the F20 shading at nearly all distances.}
    \label{fig:VPFofR_z7p7}
\end{figure}

%\textit{\textcolor{blue}{Lucia brainstorming this z=7.7 plot... maybe this entire results section needs reorganizing? let me know.}}

% As we have explored, 
To lay the framework for conceptualizing multi-redshift constraints from the VPF, let us first examine the complete VPF($R$) distribution of all our explored survey areas near just $7.6<z<7.9$. This mimics the detailed projection for the LAGER $z=6.9$ narrowband survey's VPF we carried out in Paper 1.
As Figure \ref{fig:ReionizModels} details, each of the N20, Y20, and F19 models serendipitously correspond to one of the discrete ionization fractions of the J14 volumes around $7.6<z<7.9$. We can therefore answer:
% a constraint between very late vs.\ early reionization can be made near $7.6<z<7.9$ with all the survey areas we explore. 
% Though we have so far focused on the behavior of the VPF at $R\sim12$ cMpc, we measure the VPF between $5<R<50$ cMpc. 
how do the complete VPF curves, for $5<R<40$ cMpc, behave for each of these distinct ionization scenarios near $7.6<z<7.9$?

Figure \ref{fig:VPFofR_z7p7} compares the complete VPF(R) measurements for the three models in this narrow redshift window: F19 predicting $\langle x_i \rangle _v$=0.66 near $z=7.85$; Y20 predicting $\langle x_i \rangle _v$=0.485 near $z=7.7$; and N20 predicting $\langle x_i \rangle _v<$0.22 near $z=7.6$. The dashed colored lines indicate the mean VPF across all the independent $\Delta z=0.2$ slices of a given area. The colored shaded regions indicate the (log-space) errors corresponding to the 1$\sigma$ standard deviation across the slices. The drop-off of the colored shading indicates where most slices start to find VPF($R$)=0. We also show as grey shading the maximum distance scale to measure the VPF for each survey area we probe \citep[given a precision to log$_{10}$(VPF)$>-1.5$]{Perez2021}. Since N20 predicts such a late and swift reionization, we assume the ionization fraction will be no higher than 22$\%$ near $z=7.75$, and that therefore the VPF measured for $\langle x_i \rangle _v$=0.22 near $z=7.6$ is a good approximation. Though the Roman grism becomes more sensitive past $z>7.7$, we assume that this effect would be offset by a much more neutral IGM in the N20 model. 

Much like Paper 1 showed for LAGER at $z=6.9$, we find this single redshift can yield useful constraints on reionization even with small area surveys. We find similar results for $z\sim7.7$: increasing the area of the survey yields VPF measurements that more accurately reflect the `true' VPF value of the sample with lower variance. 
Due to this, and the results of upcoming sections, 
% We note that in this work 
we advocate improving the VPF's precision by increasing the continuous covered area of a single survey region, rather than combining several independent smaller survey regions of the sky (as in Paper 1 with LAGER). With a $\sim1$ deg$^2$ survey, N20 can be well-distinguished from the Y20/F19 scenarios ($>3\sigma$), effectively ruling in or out a very late reionization history. Increasing to $\sim4$ deg$^2$ improves the constraint notably ($>6\sigma$) and also expands the distance scales that allow this constraint (to $R<30$ cMpc). However, the Y20/F19 models cannot be distinguished until analyzing $\sim15$ deg$^2$ ($\sim1.5\sigma$ at $R<15$ cMpc). 

Finally, we remind readers that we have focused the scope of this analysis to $7.6<z<7.9$ for two key reasons. First, we are limited to the ionized fractions of the existing J14 simulations. Second, we focus on the representative reionization models of F19, Y20, and N20, and when in cosmic history they predict each of the discrete fractions. However, once actually observing with Roman, we will find completely independent measurements of the VPF at more redshifts than we are able to probe under this analysis. As we have done in Paper 1 for LAGER, the exercise of this sub-section can be done at any redshift to constrain which J14 ionization fraction simulation's clustering may best describe the VPF as measured in a deep Roman LAE survey. Therefore, the overall evolution of the VPF across $7.5<z<10.5$ will help confidently constrain the pacing of our universe's reionization history. 
% \textcolor{cyan}{In the next sections, we quantify precisely how these constraints will appear for Roman deep surveys of several sizes, when focusing solely on the redshift evolution of the VPF measured for the characteristic bubble scale $R\sim12$ cMpc.} 

\begin{figure}[t]
    \centering
    \includegraphics[width=0.7\textwidth]{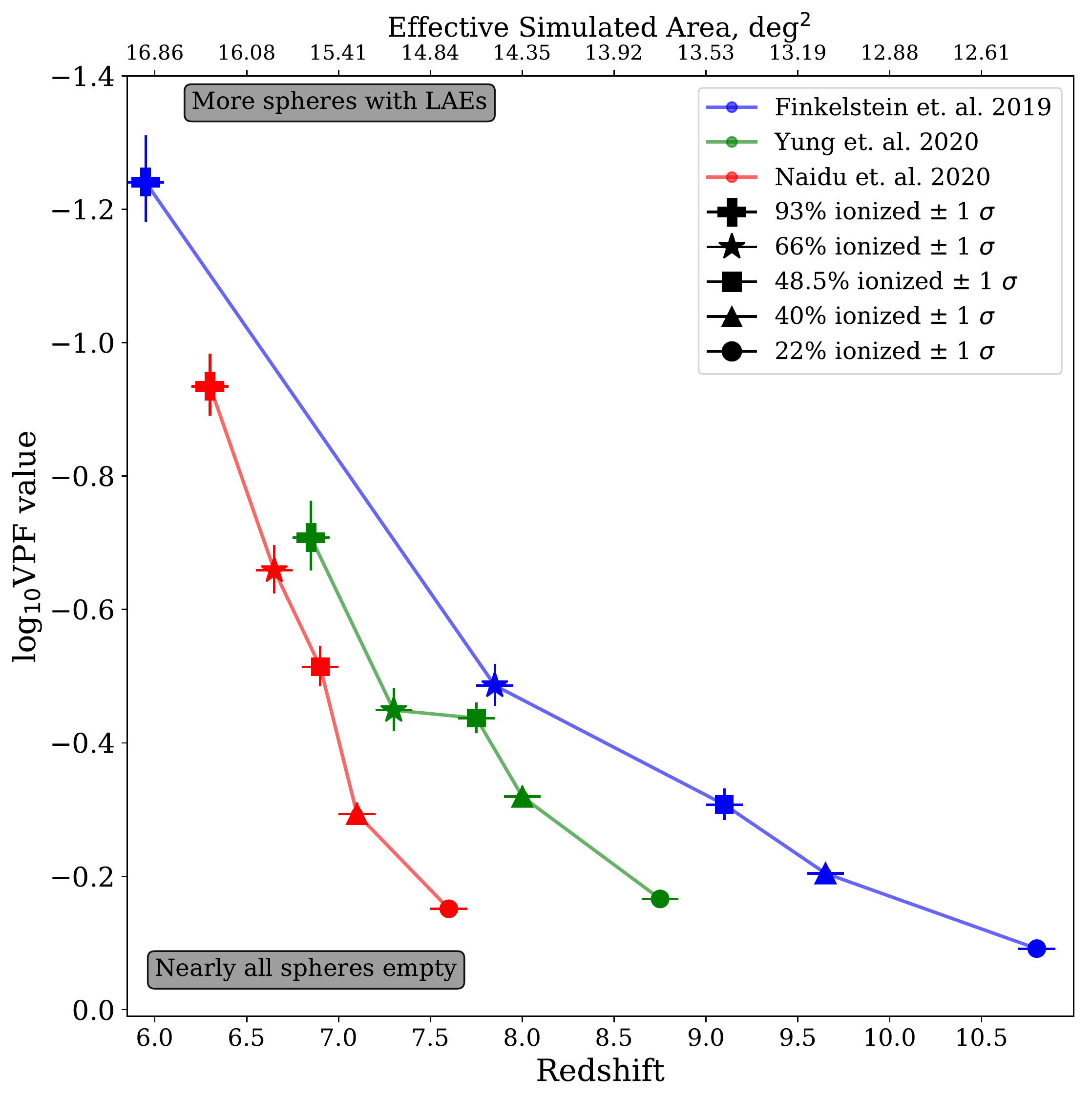}
     \caption{
     VPF($z$) for a Roman field covering 13-16 deg$^2$ at 11.86 cMpc. The colors indicate the model of reionization: N20 (red), F19 (blue), and Y20 (green). The shapes of the points indicate which J14 volume was used for the VPF measurement: $\langle x_i \rangle _v=$ 93 (plus sign, left-most), 66 (star), 48.5 (square), 40 (triangle), or 22 (circle, right-most) percent ionized. The effective simulated area in deg$^2$ corresponding to these slices, for a given redshift, is listed on the upper axis.
    %  The window histograms show the complete distribution of the VPF at this radius for all full face $\sim14-16$ deg$^2$ slices created for each simulation-reionization model pairing (at least $5\times8$ slices, each with a redshift depth of $\Delta z=0.2$).
     Error bars correspond to the 1$\sigma$ standard deviation of the VPF across all full face $\sim13-16$ deg$^2$ and $\Delta z=0.2$ slices created for each simulation-reionization model pairing. The detailed VPF distributions are shown in the Appendix. \label{fig:VPFbyZ_16deg2}}
\end{figure}

\subsection{Constraints from a 13-16 deg$^2$ survey} \label{subsec:FullFaceConstraints}

We now examine the evolution of the VPF across reionization history that Roman will observe, and identify what constraints on reionization models the VPF may yield. We focus on three representative models of reionization: the very early and slow F19 model, the quick yet middle-of-the-road Y20 model, and the very late and very fast N20 model. The complete histograms for each model's redshift-reionization scenario VPF are shown at full scale in the Appendix \ref{app:VPFdistributions} Figures \ref{fig:Finkelstein_12Mpc_histogram} - \ref{fig:Naidu_12Mpc_histogram}. We ask: how well can Roman distinguish these reionization histories with the VPF using surveys of $\sim 1$ vs.\ $\sim 4$ vs.\ 13--16 deg$^2$?

We first examine the results of the most ambitious survey we probe, for narrow slices encompassing 13-16 deg$^2$ of Roman grism observations. Figure \ref{fig:VPFbyZ_16deg2} shows the VPF($z$) at $R\sim12$ cMpc for each redshift-reionization scenario, using several independent full-face (602x607) cMpc$^2$ slices from each simulation. The 3 representative models of reionization are indicated by different colors: N20 (red), F19 (blue), and Y20 (green). The shape of each symbol indicates which J14 simulation was sliced up for the VPF measurement: 22, 40, 48.5, 66, or 93 percent ionized (marked using circles, triangles, square, stars, and crosses, respectively). The error bars are the 1$\sigma$ standard deviation across the 7-14 full-face slices' VPFs at the given radius (some of which are smaller than the symbol sizes). 
% The histograms plotted in small windows show the exact distribution of the VPF across the full-face slices, with the histograms matching in color the curve and model they correspond to. The complete histograms for each model are shown at full scale in the top row of Appendix \ref{app:VPFdistributions} Figures \ref{fig:Finkelstein_12Mpc_histogram} - \ref{fig:Naidu_12Mpc_histogram}.

These VPF($z$) curves follow some trends: more neutral and higher-redshift slices 
have higher void probabilities (i.e.\ their $\log_{10}$(VPF) is closer to 0, meaning they have more empty dropped circles). This is not unsurprising, as lower density samples implicitly have more and larger voids. However, the pattern of the VPF also incorporates the predicted reionization histories, which Roman will be able to observe simultaneously across redshift history. 

% \textit{Clarify: they never actually overlap, even the Fink73 and A58 are 2sigma apart in our pessimistic statistics.}
This largest explored survey area creates many opportunities for clear constraints on the timing and pace of reionization. First, we see the VPF is clearly different between different ionization fractions at the same redshift 
% and therefore identical luminosity selections 
(as seen in Paper 1, and further explored for $z=7.75$ in \textsection \ref{subsec:z7p75focus}). With the $\sim16$ deg$^2$ survey area, the VPF of all models' redshift-reionization scenarios are completely separated by at least 1.5$\sigma$ (e.g.\ F19 and Y20 near $z=7.8$), with most separated by $>4\sigma$ from the nearest scenario of another model.
%% Which is it?
For example, the Y20 ($\langle x_i \rangle _v$=0.93) and the N20 ($\langle x_i \rangle _v$=0.485) VPF measurements at $z=7.8$ are distinguishable to 5$\sigma$, allowing Roman to decidedly constrain between a very late vs.\ somewhat early reionization. Past $z=8$, where the Roman grism reaches peak sensitivity, the F19 ($\langle x_i \rangle _v$=0.66) and Y20 ($\langle x_i \rangle _v$=0.485) may be separated enough for distinguishing constraints of 3$\sigma$ with a survey of $\sim14$ deg$^2$. The measurement of the VPF near $z \sim 8.1$ would therefore clarify if reionization started very early or somewhat early. Additionally, the very late and fast N20 ($\langle x_i \rangle _v$=0.22) history can be distinguished from the earlier start of Y20 ($\langle x_i \rangle _v$=0.40, 0.485) or F19 ($\langle x_i \rangle _v$=0.66) to $>8\sigma$ near $z=7.75$. Finally, if the pacing and timing predicted by the F19 and Y20 VPF($z$) hold near $z\sim9$, the Roman may distinguish the two reionization histories to more than 4$\sigma$ with the VPF. 

Taken together, these results show that by combining at least two redshifts, we are able to distinguish any pair of reionization histories with the VPF of LAEs. More broadly, this VPF test will only be unable to distinguish two reionization models if they yield functionally identical reionization histories. Therefore, measurements of the Ly$\alpha$ VPF throughout the Roman grism will definitively identify which of the probed models best describes the reionization history of the universe. Additionally, we note the investment into a survey of this scale would offer other constraints on reionization, such as \lya-21cm cross-correlation (e.g.\ \citealt{Hutter2019}, who identified $\sim$20 deg$^2$ surveys to utilize SKA alongside the angular correlation function of LAEs).
% can yield a conclusive narrative for the reionization history of the universe with a large survey of 13-16 deg$^2$.} 

\begin{figure}
    \centering
    \includegraphics[width=0.7\textwidth]{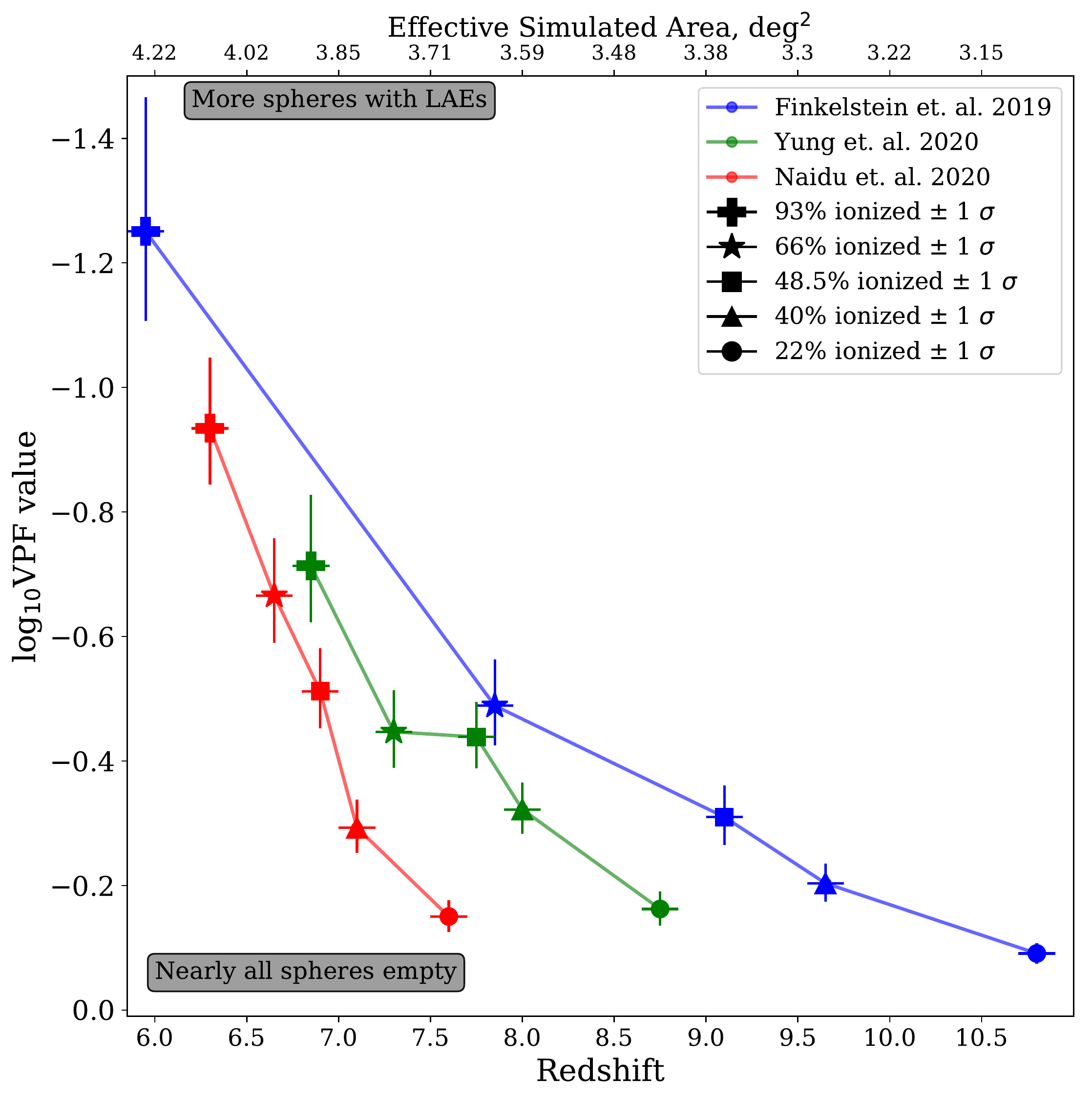}
     \caption{
     Like Figure \ref{fig:VPFbyZ_16deg2}, but instead for slices of $\sim4$ deg$^2$ created for each simulation-reionization model pairing (at least $8\times4$ slices, each with a redshift depth of $\Delta z=0.2$). 
    %  The histograms show the VPF distribution for the particular redshift-reionization scenario across all slices; the mean and 1$\sigma$ standard deviation plotted above the distribution are from the full-face slice histograms.  
  \label{fig:VPFbyZ_4deg2}}
\end{figure}

\begin{figure}
	\begin{center}
    \includegraphics[width=.75\textwidth]{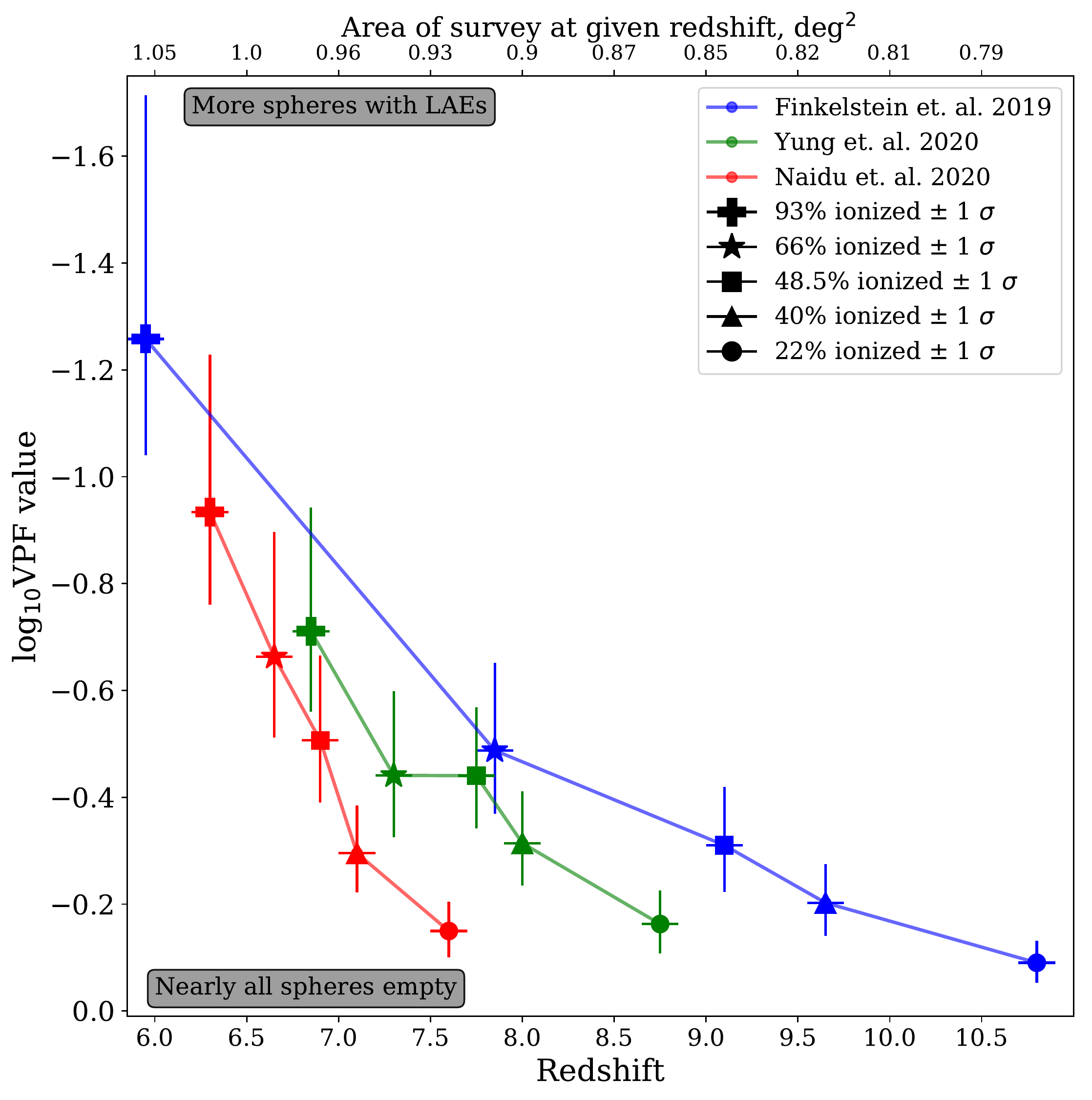}
	\caption{ 
	Like Figures \ref{fig:VPFbyZ_16deg2} and \ref{fig:VPFbyZ_4deg2}, but instead for slices of approximately 1 deg$^2$ (one sixteenth of the face of the \citealt{Jensen2014} simulations; at least 16$\times$8 total slices). Error bars are the 1$\sigma$ standard deviation across all slices' VPF. %A survey covering 1 deg$^2$ near $z\sim7.7$ could distinguish very late vs. early reionization to approximately 3$\sigma$ with the VPF.
	\label{fig:VPFbyZ_1deg2}}
	\end{center}
\end{figure}

\subsection{Constraints from 4 and 1 deg$^2$ surveys} \label{subsec:SmallerAreaConstraints}

A LAE survey of 13-16 deg$^2$ with Roman would yield excellent $>5\sigma$ constraints at several redshifts, and decisively rule in or out key features of reionization history. Might we reach similar constraints with smaller surveys? We repeat the process of Figure \ref{fig:VPFbyZ_4deg2} using instead the $\sim 4$ deg$^2$ and $\sim 1$ deg$^2$ slices. 

The primary effect of the smaller area is decreased precision in the VPF from Poisson noise and larger variance across slices due to the decreasing total number of LAEs. However, we find the VPF across the $\sim 4$ deg$^2$ slices can still reach multiple constraints for the epoch and pace of reionization, to between 2.5 - 8$\sigma$ across $6.8<z<9$. 
% , such as mostly neutral vs. mostly ionized near $z=8$ and $z=9$. 
First, the Y20 ($\langle x_i \rangle _v$=0.93) and the N20 ($\langle x_i \rangle _v$=0.485) VPF measurements at $z\sim6.8$ are distinguishable to about 2.5$\sigma$. This would identify a very late reionization at redshifts accessible with the Roman prism and ground-based surveys (consistent with our results in Paper 1 for LAGER). 
Additionally, near $z=7.5-8$ the F19 ($\langle x_i \rangle _v$=0.66) and N20 ($\langle x_i \rangle _v$=0.22) scenarios could be distinguished to 3$\sigma$ apart, independently constraining early vs.\ late and slow vs.\ fast reionization.
Notably, more constraints for the timing of reionization are distinguishable to $6-8\sigma$ near $z=7.75$ between the N20 ($\langle x_i \rangle _v$=0.22) and Y20 ($\langle x_i \rangle _v$=0.40, 0.485) models.
However, we note these constraints are within the redshift range where the sensitivity of the grism changes rapidly, so LAE selection must be done carefully to leverage this survey area.
% , and \citet{Naidu2020} ($\langle x_i \rangle _v$=0.485) and \citet{Yung2020} ($\langle x_i \rangle _v$=0.93) are distinguishable to 2-3$\sigma$.
% \textcolor{blue}{Once get 95: list here! should be quite noticeable, but it goes VERY low z for Fink, not super comparable... ok Yung 95 VPF 2-3sigma from Naidu58 near z=7}
Finally, the F19 ($\langle x_i \rangle _v$=0.485) and Y20 ($\langle x_i \rangle _v$=0.22) scenarios at $z\sim9$ may be at least $3-4\sigma$ apart if the apparent patterns hold beyond where the discrete fractions of the J14 simulations reach.

Is it possible to make constraints with an even smaller survey? Figure \ref{fig:VPFbyZ_1deg2} repeats our process for the $\sim1$ deg$^2$ slices at $R=11.86$ cMpc. The reduced area puts some of our earlier constraints out of reach, as fewer galaxies introduce more uncertainty to the VPF and makes fractions more difficult to distinguish. This scarcity of LAEs can still be a useful constraint: for example, an early investment of a few weeks of Roman observing time to cover 1 deg$^2$ may find fewer than 50 LAEs deg$^{-2}$ at $z>8$, which would provide some support for the later reionization models (i.e. measure a low surface density consistent with the more neutral ionization fractions). However, there are still strong constraints the VPF of LAEs will yield for the timing of reionization with a $\sim1$ deg$^2$ survey. The N20 ($\langle x_i \rangle _v$=0.22) and Y20 ($\langle x_i \rangle _v$=0.40, 0.485) histories near $z=7.75$ are still distinguishable to $3-4\sigma$ (assuming LAE selection that accounts for the rapidly changing grism sensitivity to \lya). Additionally, the scenarios of F19 (near $\langle x_i \rangle _v$=0.485) and Y20 ($\langle x_i \rangle _v$=0.22)  at $z\sim9$ could perhaps be distinguished to more than $2\sigma$ apart, if the apparent trend of we see in the VPF($z$) holds.

\section{Conclusion} \label{sec:Conclusion}

% \subsection{Considerations of this work}
% \textbf{Ly$\alpha$ duty cycle:} Thing to explore: how degenerate is this with duty cycle? if do 10 vs 15$\%$, for example? Disclaimer for tables: take these as rough guides, model-dependent and duty-cycle dependent! If data on Ly$\alpha$ duty cycle changes, number densities will change, clustering will shift. How much might they change? \textit{try with 10 percent duty cycle!!!} First 3 models appear to drop 5-8 percent with DC change (compared to 12.5, the present DC that Lucia assumes); very neutral fractions lose about 15-20$\%$.
% Perhaps the take-away is to get a little extra area, just in case
% Thinking about \textbf{cosmic variance}... the 1deg2 should definitely see the differences, but maybe not 4deg2, try the calculator just in case 
% look at anne hutter's paper: with 2ptCF, need 20deg2!!!! THIS is our pinnacle result! the hype is real!!!!

This work explores the constraints the Roman Space Telescope will find for the timing and pace of reionization based on the clustering of LAEs. 
% How large of a survey must Roman observe to distinguish between early and late reionization? Or fast and slow reionization? 
We use the Void Probability Function (VPF), which asks how many randomly dropped spheres are empty, is tied to all higher order correlation functions, and is a simple clustering statistic to implement and guide survey construction. We focus on three representative models for reionization: \citet[very early and slow reionization]{Finkelstein2019}, \citet[moderately early and quick reionization]{Yung2020}, and \citet[late and very fast reionization]{Naidu2020}. 
% With the VPF, we refine what Roman LAE surveys are required to distinguish key models for the reionization history of the universe. 
We analyze ($602\times607\times600$) cMpc$^3$ simulations of LAEs through reionization at discrete ionized IGM fractions between $0.22 < \langle x_i \rangle_v < 0.93$ \citep{Jensen2014}. Informed by detailed simulations of  Roman's grism responsiveness to Ly$\alpha$  (Wold et al.\ in prep), we create flux-limited mock LAE samples as may be observed by Roman throughout reionization.

We mimic the three models' reionization histories by redshifting each J14 simulation to the redshifts where each model predicts the given volume-averaged ionization fraction. We create mock Roman LAE surveys of $\Delta z=0.2$ that cover: 13-16 deg$^2$ (the full face of the simulation), $\sim4$ deg$^2$ (quarter face of the simulation), or $\sim1$ deg$^2$ (one sixteenth of the simulation face). We measure the VPF and answer: what constraints on the pacing and timing of reionization might Roman find with the clustering of LAEs, as a function of survey area?

% We measure the VPF across all mock surveys for all redshift-reionization scenarios, and compile a centralized narrative of how LAE clustering will be observed and constrained by Roman. We focus on the VPF measured specifically at $R\sim12$ and $R\sim28$ cMpc. 
We find that a $\sim1$ deg$^2$ survey can distinguish between very late vs.\ early reionization near $z=7.7$ to $>3\sigma$, and between very early and slow vs.\ quick and late reionization near $z\sim9$ to $>2\sigma$.
% can give an initial indication of whether late or early reionization is preferred by how many LAEs are found with the Roman grism, but find great payoff in constraining power with slightly more invested survey area. 
By investing in $\sim4$ deg$^2$, the VPF of LAEs is able to distinguish: early vs.\ late reionization to $3\sigma$ between $7.5<z<8$ and $z\sim9$; and slow vs.\ fast reionization to $3\sigma$ at $z\sim8$. Additionally, early and slow vs.\ fast and late reionization may be distinguished to $6\sigma$ or more near $z=7.5-8$.
% we find the most promising clustering analysis to distinguish these models of reionization will be at $z\sim8$ and $z\sim9$. 
% Though the Roman grism is still ramping up at $z\sim7.5$,
Finally, we find a 13-16 deg$^2$ area would give VPF measurements that would essentially trace out a precise reionization history of the universe, and determine with great confidence ($>4-5\sigma$) the most accurate model describing the timing and pace of reionization. 
% distinguish \citet{Finkelstein2019}, \citet{Yung2020}, and \citet{Naidu2020} from each other to great precisio. 
% An additional measurement of the VPF at $z\sim8.5$--where Roman is most sensitive, and therefore would require as few as 2-4 deg$^2$--would offer an independent distinguishing constraint between \citet{Finkelstein2019} and \citet{Yung2020}. 

\begin{acknowledgments}
% We thank Hannes Jensen and Garrelt Mellema for sharing the simulations that this work analyzed. We also thank Garrelt Mellema for comments offered that improved the manuscript. We thank the larger LAGER collaboration for their support of this work. We thank the anonymous reviewer for their comments that improved the clarity and context of this manuscript. Finally, we are grateful for the continual support from Isak Wold as we refined our experiment setup using \citet{Wold2022}. 
This work was supported by NSF grant AST-1518057, as well as NASA support to A.S.U. under contract NNG16PJ33C, “Studying Cosmic Dawn with WFIRST.” LAP is also supported by the Princeton Future Faculty in the Physical Sciences Postdoctoral Fellowship.
IGBW is supported by an appointment to the NASA Postdoctoral Program at the Goddard Space Flight Center. The material is based upon work supported by NASA under award number 80GSFC21M0002.
% The Cosmic Dawn Center (DAWN) is funded by the Danish National Research Foundation under grant No. 140. 
% Part of this work was carried out on the \textit{Saguaro} supercomputing cluster operated by the Fulton School of Engineering at Arizona State University, and also 
This work made use of NASA's Astrophysics Data System Bibliographic Services. Our figures were made with \textit{Matplotlib} \citep{Matplotlib}, and the bulk of our calculations used \textit{Numpy} \citep{Numpy2020}, \textit{Scipy} \citep{Scipy2020}, and the \citet{NedWright} cosmology calculator.

\end{acknowledgments}

\newpage 

\appendix

% \section{TO NOT PUBLISH: Nuance of Jensen Simulated LAEs}

% \textit{Comparing the transmitted log$_{10}$(L$_{\textrm{Ly}\alpha}$) distributions of each \cite{Jensen2014} simulation; the dashed lines are the mean of each distribution. And hopefully answer why the 83$\%$ simulation is a little weird--it has a mean luminosity a good bit lower than 73$\%$, meaning that it will lose more on average under a similar flux cut... which is what we see! Maybe this is why it might be extra prudent to downsample to one number density, oops. Common sense reminder: NONE of the ionization models predict anything more ionized than 73 at the redshifts the RST grism can observe! We literally won't work with 83, 92, and 95.}

% more explanation for weird 58 and 83; all the simulations have slightly different shapes of line luminosity distributions, may have a big effect and the high tail we probe.

% % \begin{figure}
% % 	\begin{center}
% % \includegraphics[width=.75\textwidth]{CommonSense_checkLyAluminosities.png}
% % 	\caption{Comparing the transmitted log$_{10}$(L$_{\textrm{Ly}\alpha}$) distributions of each \cite{Jensen2014} simulation; the dashed lines are the mean of each distribution.}
% % 	\label{fig:CheckLs}
% % 	\end{center}
% % \end{figure}

% % \begin{sidewaysfigure}
% \begin{figure}
% 	\begin{center}
% \includegraphics[width=\textwidth]{LuminosityFunctions_JensenSimsRaw.pdf}
% 	\caption{Differential luminosity functions for the completely unaltered J14 LAE catalogs we work with.}
% 	\label{fig:rawLFs}
% 	\end{center}
% \end{figure}
% % \end{sidewaysfigure}

\section{Detailed VPF distributions} \label{app:VPFdistributions}

\begin{sidewaysfigure}
    \centering
    \includegraphics[width=0.85\textwidth]{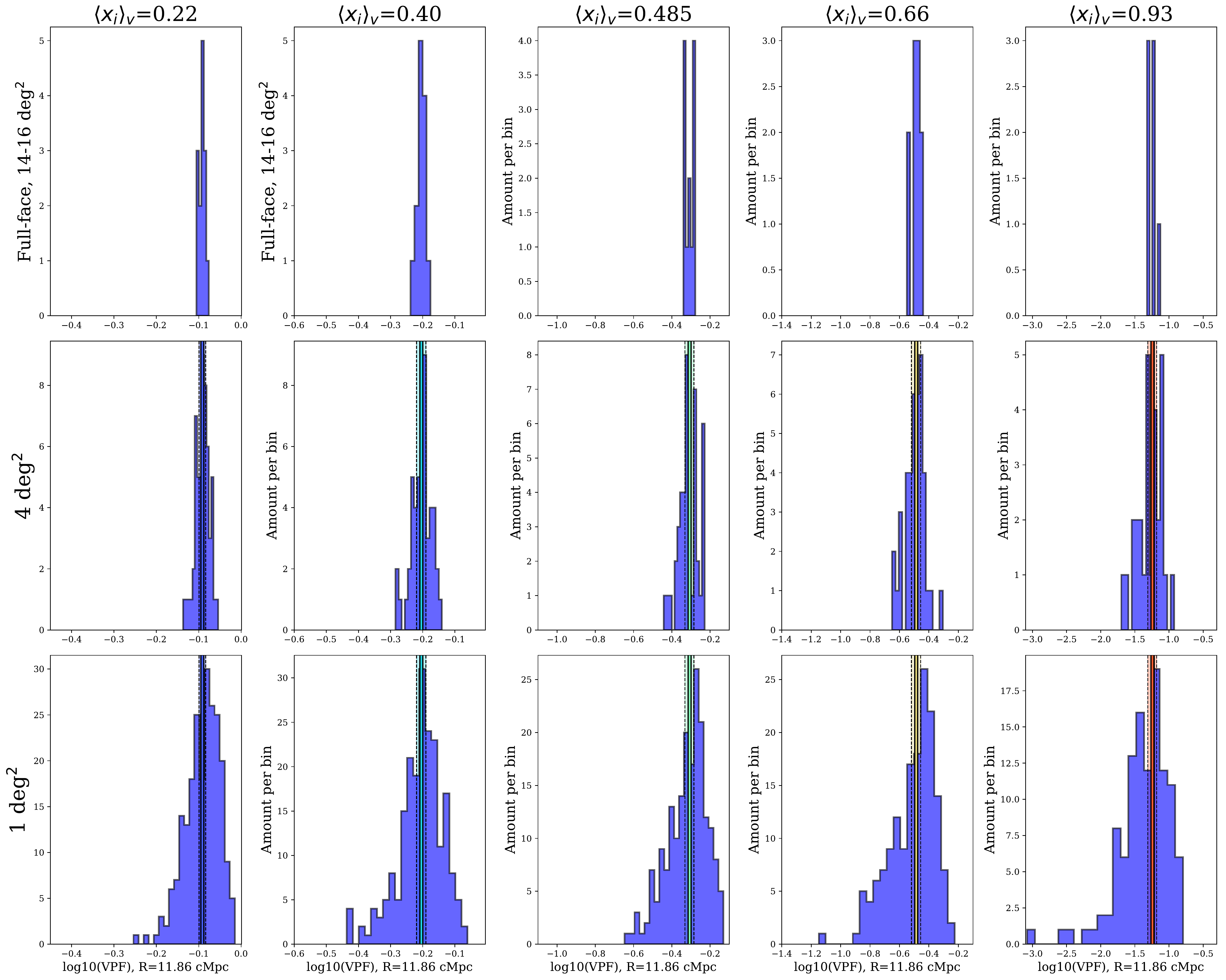}
	\caption{
	Focused distributions of the VPF measured at $R=11.86$ cMpc for the \citet{Finkelstein2019} redshift-reionization scenarios; the columns are the J14 simulation volume (left: $\langle x_i \rangle_v$=0.22, in increasing order until right: $\langle x_i \rangle$=0.93). The rows are for the Roman survey area probed: full-face 13-16 deg$^2$ (top), $\sim4$ deg$^2$ (middle), $\sim1$ deg$^2$ (bottom). The solid lines in all figures indicate the mean VPF across the full-face slices; the semi-transparent shading in the $\sim4$ deg$^2$ and $\sim1$ deg$^2$ histograms are the 1$\sigma$ standard deviation measured on the full-face distributions. 
% 	Histograms with grey backgrounds need a larger survey area for a trustworthy VPF.
% 	The colors of these lines and shaded regions correspond to the color of the ionization fraction dashed lines in Figure \ref{fig:ReionizModels}. 
    \label{fig:Finkelstein_12Mpc_histogram}}
\end{sidewaysfigure}

\begin{sidewaysfigure}
    \centering
    \includegraphics[width=0.9\textwidth]{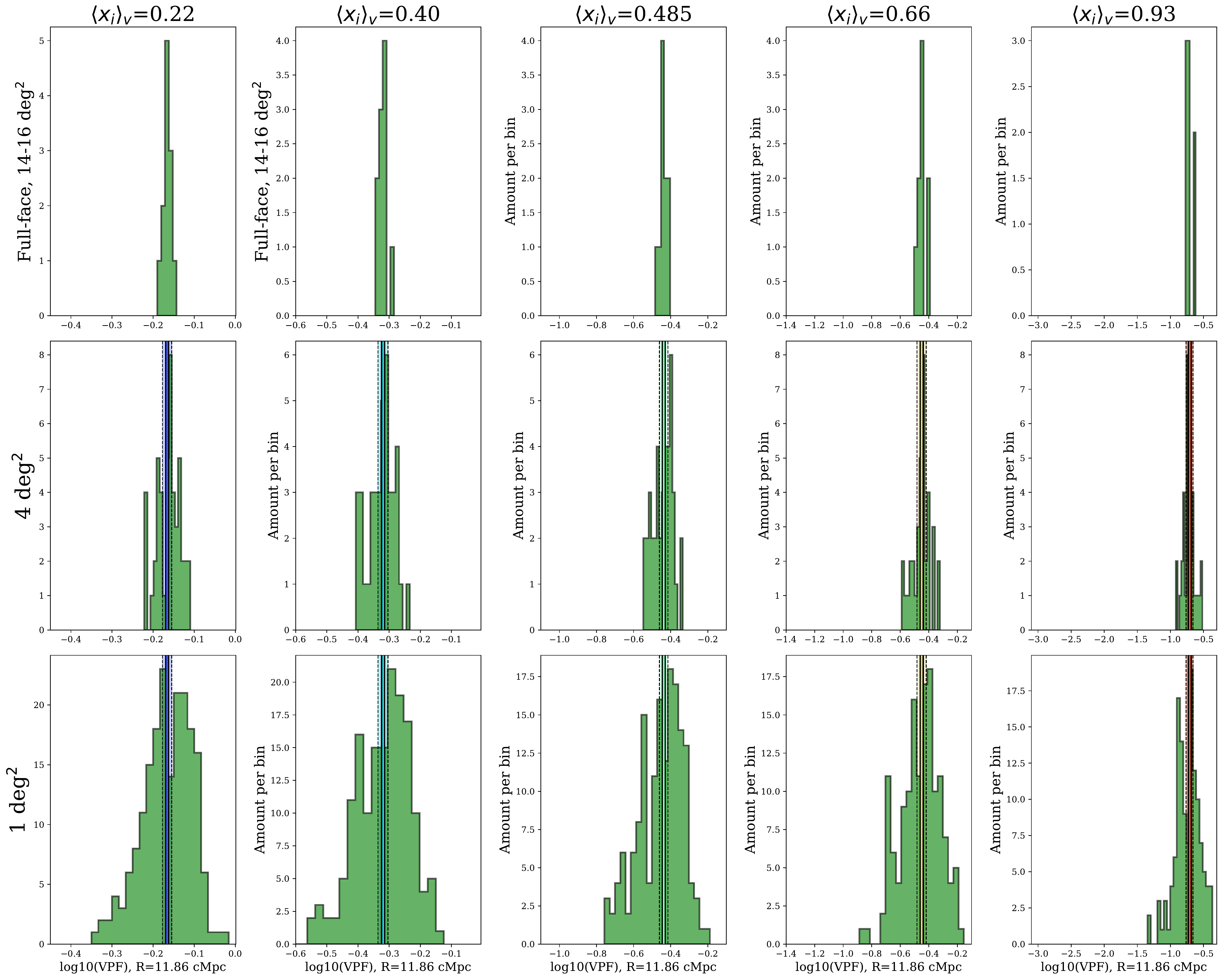}
	\caption{
	Like Figure \ref{fig:Finkelstein_12Mpc_histogram}, but for \citet{Yung2020}. \label{fig:Aaron_12Mpc_histogram} }
\end{sidewaysfigure}

\begin{sidewaysfigure}
% \begin{figure}
    \centering
    \includegraphics[width=0.9\textwidth]{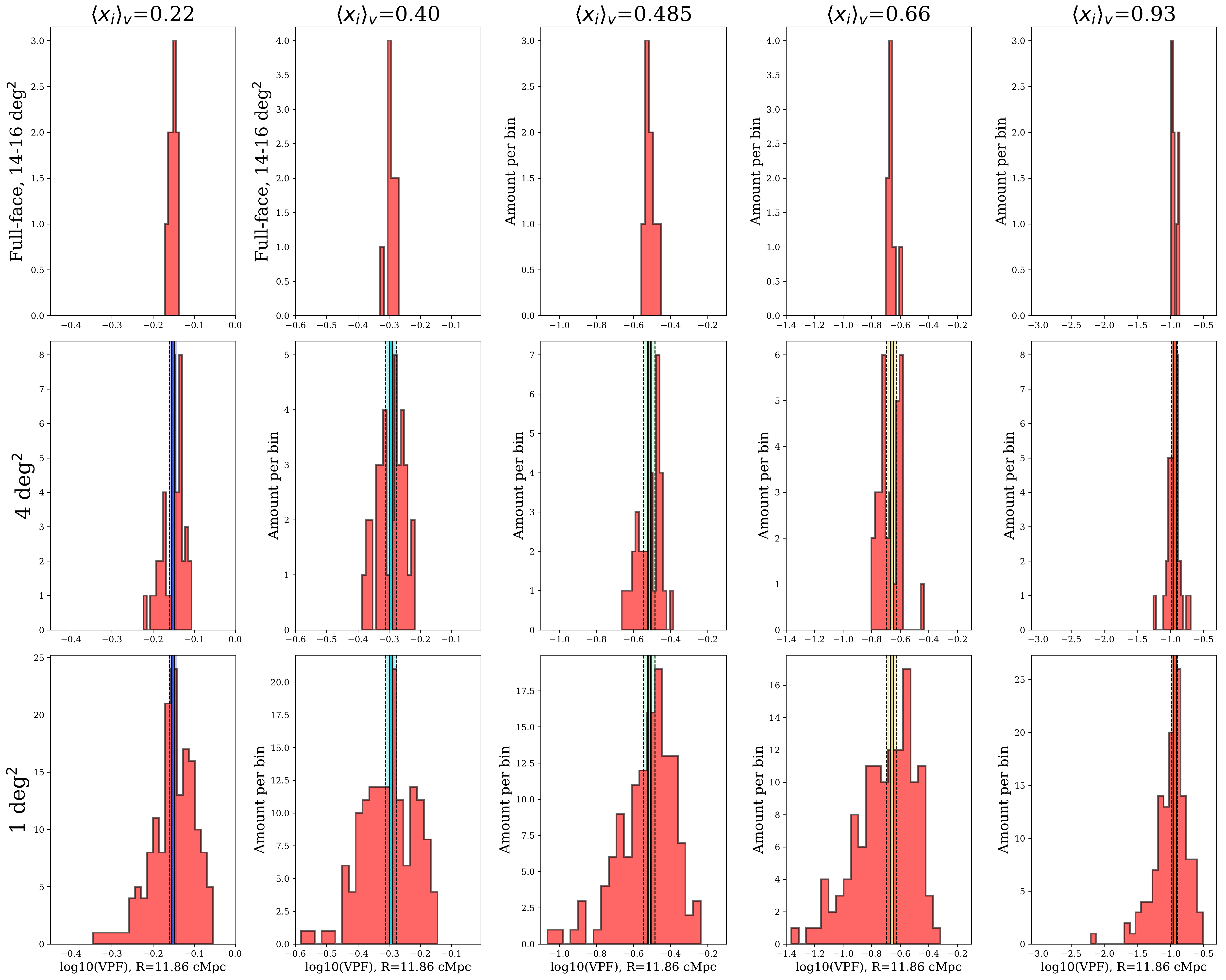}
	\caption{
	Like Figure \ref{fig:Finkelstein_12Mpc_histogram}, but for \citet{Naidu2020}.  \label{fig:Naidu_12Mpc_histogram}}
% \end{figure}
\end{sidewaysfigure}

What do the distributions of the VPF across all the independent $\Delta z=0.2$ slices look like for each redshift-reionization scenario? We compare the distribution of the VPF measurements across: the `full-face' slices of approximately 13-16 deg$^2$ (top row), the few dozen slices of approximately 4 deg$^2$ (middle row), and the several dozen slices of approximately 1 deg$^2$ (bottom row) in Figures \ref{fig:Finkelstein_12Mpc_histogram} - \ref{fig:Naidu_12Mpc_histogram}. Columns indicate the J14 volume analyzed in the scenario, with the most neutral to the left and increasing in ionization fraction. 
% Histograms with grey shaded backgrounds indicate redshift-reionization combinations for which the survey area is not large enough for a trustworthy VPF measurement. 
Each figure shows, in a solid line whose color corresponds to the $\langle x_i \rangle _v$ value of the simulation (e.g. in Figure \ref{fig:ReionizModels}, the mean of the VPF across the several full-face slices). The histograms for the 4 deg$^2$ and 1 deg$^2$ slices also show the 1$\sigma$ standard deviation across the full face slices. Increasing the survey area steeply narrows the distribution of sampled slices about what is likely the true VPF value.

The histograms are not normalized, and we use (5, 15, 20) default-\textsc{Matplotlib} bins for the ($\sim16$, $\sim4$, $\sim1$) deg$^2$ slices. 
% \textit{We measured the VPF of five re-sampled full-face slices (at the duty cycle step, we re-apply it five times for five probably unique samples of LAEs), and of twice-resampled 4 deg$^2$ slices, in order to improve the number statistics in generating these histograms.}
% We note that the $x$-axis for the log$_{10}$(VPF) values are kept steady across Figures--this is so that the histograms can be directly compared to at a glance model-by-model in Figures \ref{fig:VPFbyZ_16deg2} - \ref{fig:VPFbyZ_11p86_1deg2}. 
We also remind readers that each slices' VPF measurement is the average across 5 calculations of the VPF with the \citet{Banerjee2021} $k$-NN method. Though not shown here, all valid distributions are clustered compared to random (e.g.\ Paper 1 Figures). 
% Additionally, slices with no LAEs are assigned a false VPF value of -6 (off the scale), leading to very large 1$\sigma$ standard deviations at e.g. \citet{Finkelstein2019} $z=10.8$ and $\langle x_i \rangle _v =0.22$ scenario for 1 deg$^2$. 
Finally, a quirk of the VPF is that the spread of a distribution is related to its central value: more clustered VPFs (closer to zero) also have narrower distributions. 

\clearpage

\bibliography{sample631}{}
\bibliographystyle{aasjournal}

\end{document}